# VNF and Container Placement: Recent Advances and Future Trends

Wissal Attaoui, *Student Member, IEEE*, Essaid Sabir, *Senior Member, IEEE*, Halima Elbiaze, *Senior Member, IEEE*, Mohsen Guizani, *Fellow Member, IEEE*

Emails: w.attaoui@ensem.ac.ma, e.sabir@ensem.ac.ma, elbiaze.halima@uqam.ca, mguizani@ieee.org

*Abstract*—With the growing demand for openness, scalability, and granularity, mobile network function virtualization (NFV) has emerged as a key enabler for most mobile network operators. NFV decouples network functions from hardware devices. This decoupling allows network services, referred to as Virtualized Network Functions (VNFs), to be hosted on commodity hardware which simplifies and enhances service deployment and management for providers, improves flexibility, and leads to efficient and scalable resource usage, and lower costs. The proper placement of VNFs in the hosting infrastructures is one of the main technical challenges. This placement significantly influences the network's performance, reliability, and operating costs. The VNF placement is NP-Hard. Hence, there is a need for placement methods that can scale with the issue's complexity and find appropriate solutions in a reasonable duration. The primary purpose of this study is to provide a taxonomy of optimization techniques used to tackle the VNF placement problems. We classify the studied papers based on performance metrics, methods, algorithms, and environment. Virtualization is not limited to simply replacing physical machines with virtual machines or VNFs, but may also include micro-services, containers, and cloud-native systems. In this context, the second part of our article focuses on the placement of containers in edge/fog computing. Many issues have been considered such as traffic congestion, resource utilization, energy consumption, performance degradation, security, etc. For each matter, various solutions are proposed through different surveys and research papers in which each one addresses the placement problem in a specific manner by suggesting single objective or multi-objective methods based on different types of algorithms such as heuristic, meta-heuristic, and machine learning algorithms.

*Index Terms*—Virtual network function; Container; placement; 5G network slicing; Cloud Native.

## I. INTRODUCTION

### A. Motivation and New Trends

Nowadays, with the tremendous growth of mobile service demands, operators have to provide low latencies and high throughput, hence the need to rethink traditional physical architectures. Therefore, 5G tend to be virtualized; eventually, the 5G core network, i.e., the set of transmission and switching media in which the most crucial part of the traffic is processed, will no longer be carried by physical equipment as in 4G architectures, but it will be supported by the software. This virtualization is based on software-defined networking (SDN), and network function virtualization (NFV) technologies [7]. It offers many advantages and helps to provide personalized connectivity services through network slicing technology. Thus, the mobile operators will be able to agilely activate the functions required for each service, adapting the network sizing and topology according to customer needs and cloud properties: low or high throughput, low latency, high reliability, more or less distributed architecture, etc.

Mobile operators are currently experiencing a surge in 5G adoption, prompting service providers to implement the most recent Stand Alone (SA) 5G Core (5GC) [111]. Like any invention, many years of design and redesign were involved in making the 5G vision a reality, and the development process is still ongoing today. Consumers may now have access to 5G technology, but the backroom work and rework continues.

One of the issues that 5G has to deal with is figuring out how to effectively use infrastructure as a service to furnish flexibility, security, dependability, and, eventually, profitability. What began as an experiment in leveraging NFV to run VMs on hardware has swiftly evolved into VMs on shared computing and storage farms (Cloud). An orchestration layer was crucial to reduce operational overheads, but resource utilization with VMs remains a sticking point and switching to a Container architecture (CNFs) became necessary. The architecture of the 5G mobile network control plane can be a hybrid architecture between cloud-native applications and virtualization [140]. As shown in Figure 1, the network virtualization approach transforms traditional network appliances with non-standard hardware into software-based virtual machines installed in standard equipment. Network functionalities that were previously developed as monolithic programs are now split down into smaller micro-services and delivered as containers in both public, and private clouds using the cloud-native method [21]. These micro-services containers are orchestrated and supplied automatically using Continuous Integration and Deployment (CI/CD). Smaller micro-services are currently being provided by independent software suppliers who used to provide full-fledged network operations.

The 5G deployment will be gradual (i.e., initially, 4G and 5G will coexist seamlessly, as was the case for 3G and 4G). In addition, pooling physical infrastructure through virtualization techniques opens the way to create a universal 5G network core that is agnostic to the type of access (i.e., wireless,

W. Attaoui is with NEST Research Group, ENSEM, Hassan II University of Casablanca, Morocco.

E. Sabir is with NEST Research Group, ENSEM, Hassan II University of Casablanca, Morocco, and with the Department of Computer Science, University of Quebec at Montreal (UQAM), Montréal, QC H2L 2C4, Canada.

H. Elbiaze is with the Department of Computer Science, University of Quebec at Montreal (UQAM), Montréal, QC H2L 2C4, Canada.

M. Guizani is with Qatar University, Doha, Qatar.

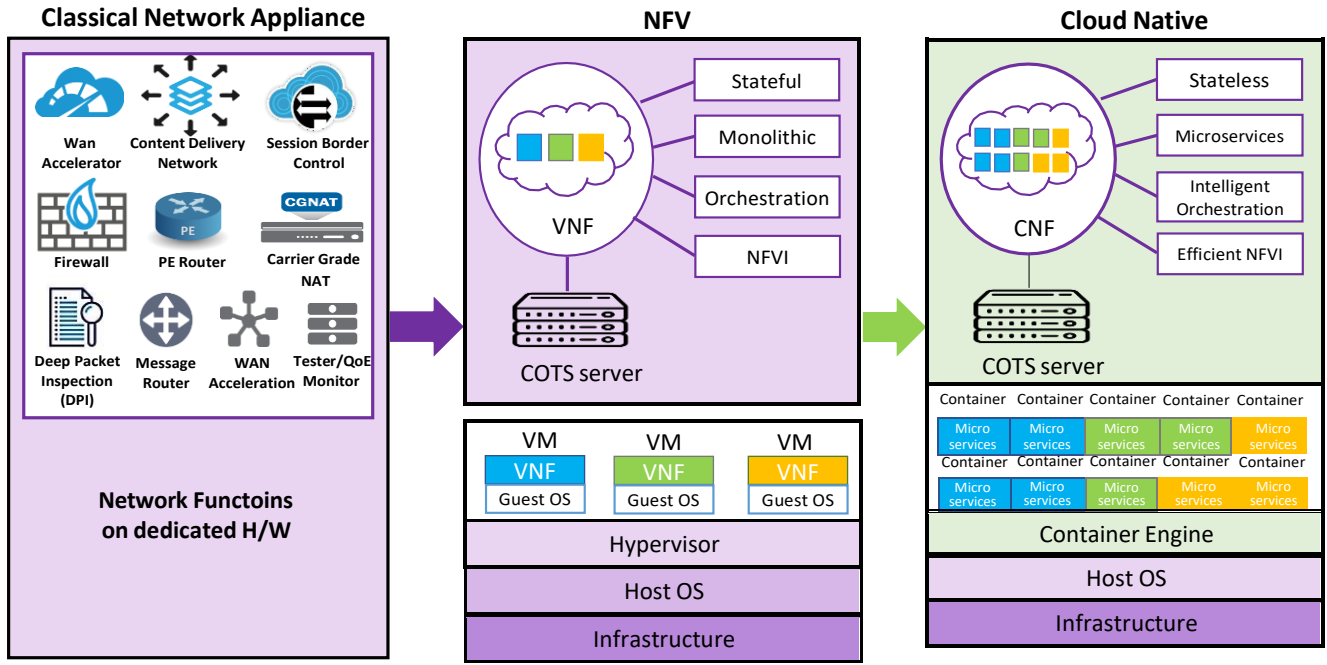

Figure 1: Monolithic architecture, VNF virtualization architecture and CNF Cloud architecture

wireline, etc.). Therefore, this will ensure homogeneous management of the operator's network. In this way, network slicing will enable mobile network operators to manage different virtual networks on the same physical network infrastructure [147]. The "slices" have features adapted in real-time to the users' needs (i.e., capacity, latency, reliability, etc.) whereby three new slicing services will be supported by 5G:

· Massive Machine Type Communications (mMTC): present the communications between many objects with varying quality of service (QoS) requirements to match the exponential increase of connected objects density;

· Enhanced Mobile Broadband (eMBB): is related to ultra-high-speed outdoor and indoor connection with uniform quality of service, even at the edge of the cell;

· Ultra-reliable and Low Latency Communications (uRLLC): ultra-reliable communications are used for critical needs with very low latency and increased responsiveness.

5G communication, with its enhanced characteristics as high bandwidth and low latency, is ideally positioned to meet the expectations of smart cities. According to predictions, cities will house half of the world's population by 2050 [50], resulting in billions of Internet of Things (IoT) devices. There will be two issues in this situation. On the one hand, smart IoT services' real-time nature is seriously compromised. Massive numbers of smart IoT devices receiving data packets cause congestion in the central cloud which may degrade the QoS. On the other hand, inflexibility in computing resource allocation is highlighted. It is challenging to allocate computing resources to smart IoT devices as they have various characteristics.

Including virtualization in 5G will help to reduce latency, provide high speed, increase scalability and improve energy efficiency. However, the real problem relies on how to place a VM or Virtual Network Function (VNF) in a cloud infrastructure optimally.

Virtualization provides the flexibility to quickly move a VM from a specific host to another without turning it off. Therefore, it can provide dynamism on VM placement with a marginal performance impact [59]. A virtual instance can be added or deleted at any time. Despite its considerable advantages, this dynamic can lead to sub-optimal or volatile configurations of virtual networks. In previous research (before 2010), the cloud controllers manage the VM placement. However, current studies are increasingly focusing on the native dynamics of VM placement as each VM has its lifetime and can experience load changes during its life cycle. Therefore, it is necessary to define the VM placement requirements, know the status of each instance and correctly determine the essential constraints, required to guarantee high performance. Operating system-level virtualization based on containers is a relatively modern technique of virtualization. Containers use the host operating system and do not require a separate one for each container resulting in lower hardware requirements than VMs. Orchestrators are required for managing and defining rules and constraints of container placement, and performance [35].

### B. On-demand computing in 5G

During the last decade, the cloud has evolved into a successful computing paradigm for delivering on-demand services over the Internet. The cloud data centers adopted virtualization technology to efficiently manage resources and services. Advances in server virtualization contribute to the cost-efficient management of computing resources in the cloud data centers. Cloud computing allows consumers to use on-demand computing resources in the form of instances (VMs or containers) rather than building physical infrastructure. These resources can be quickly delivered and handled effortlessly by cloud computing providers. It offers many interesting benefits for manufacturers and end-users, such as on-demand virtual resource provisioning, self-service capability, resource pooling, high elasticity, flexibility, and scalability. In addition, edge and fog computing are coined to complement the remote cloud to meet the service demands of a geographically distributed large number of IoT devices.

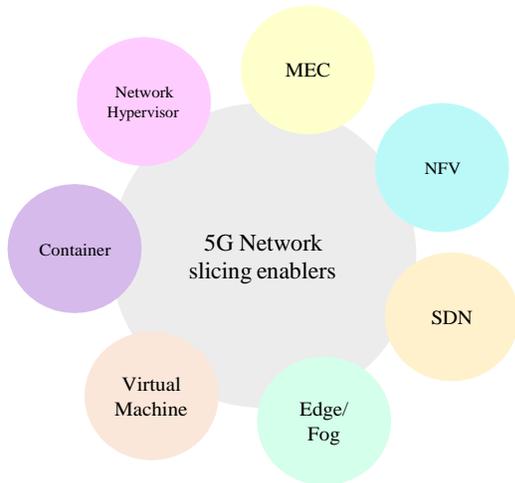

Figure 2: 5G Network Slicing enablers

5G requires a complete makeover compared to previous generations due to exigent requirements and fast-changing new use-cases. The slicing-based one-network-fits-all strategy must meet these complex and ambitious goals. Network slicing (NS) enables service providers to construct and configure their networking infrastructure to meet their own needs and tailor it for various complex scenarios. Cloud computing is likely an inextricable aspect of 5G services, serving as a superior backend for apps running on accessing devices. In this way, VMs and containers may execute VNFs in a chained configuration to offer a flexible 5G network service or application, laying the foundation for 5G network slicing. Figure 2 shows the 5G network slicing enablers, including SDN, NFV, Mobile Edge Computing (MEC), cloud/Fog computing, network hypervisors, virtual machines, and containers. Despite the numerous advantages and dynamic nature of VNF placement to create 5G network slices, if the placement was not chosen carefully, this may lead to sub-optimal or unreliable results.

Therefore, many issues should be considered in VM, VNF, and container placement (e.g., power consumption, traffic,

resource wastage, security, QoS, cost, etc.). For each matter, various solutions are proposed through different surveys and research papers in which each one addresses the placement problem in a specific manner by suggesting various methods based on different types of algorithms such as heuristic, meta-heuristic, deterministic, and AI learning algorithms.

## C. Our Contribution

The scope of this survey includes all research results on the efficient placement of any computational resources (i.e., VNF, containers) over various scenarios that differ in the environment, providing constraints and parameters required to enhance the performance and the QoS of the hosted applications. This paper presents a detailed overview of virtual resource placement in evolving cloud infrastructures handled by mobile/fixed network operators and cloud providers. This article examines the related literature over the period 2016-2021. The contributions of this survey are summarized as follows:

· We present an overview of 5G network slicing and VNF placement challenges raised in the literature, and we provide a summary of proposed solutions.
· We classify the proposed solutions according to the objective functions and the adopted techniques.
· We discuss the recent advances for 5G and the convergence toward container placement.

## D. Related surveys

As summarized in Table I, some previous surveys have already explored the field of VM and VNF placement in cloud computing. Frederico et al. [117] present a literature review of VNF forwarding graph embedding (VNF-FGE) where the classification of VNF placement solutions is based on whether adopting online or offline approaches and the algorithms are categorized into exact, heuristic, and meta-heuristic. Abdulaziz et al. [11] provide a detailed review of VM placement approaches by classifying the solutions based on adopted algorithms or models (i.e., heuristics, meta-heuristics, matching, and Multi-Criteria Decision Making (MCDM)). Xin et al. [82] discuss the network function placement and orchestration frameworks. Sedef et al. [40] propose a taxonomy of VNF placement solutions where optimization methods are divided into four types: linear programming, non-linear programming, heuristic algorithms, and Machine Learning (ML) algorithms. However, they focus only on cost, energy, and latency minimization. Omogbai Oleghe [99] highlights the concept of container placement and migration in edge computing. He studied the scheduling problem from the provider perspective by listing the frameworks and algorithms used to model and solve the container placement issues. Adel et al. [28] explore research proposals for network slice orchestration in various platforms, including VM/VNF placement in the cloud, fog, and edge computing. Balázs et al. [127] introduce a comprehensive survey of computational units placement strategies in edge infrastructures; the papers were classified based on mathematical models, objective functions, and application structure. To the best of our knowledge, this paper is the first one handling the container and VNF placement and highlighting the importance of machine learning algorithms to solve the complex problems of virtual resource orchestration such as multi-dimensional and dynamic workload characterization and auto-scaling.

### E. Article Structure

This paper is organized as follows. Section II presents the basic concepts related to cloud computing and virtualization. This paper is divided into two main parts. The first one, presented in section III, addresses the problems of VNFs placement and presents a classification of the proposed solutions based on their objective functions and algorithms. Similarly, section IV presents a new perspective towards cloud-native by handling the placement of containers in the cloud, edge, and fog computing. Section V delivers some concluding remarks and future directions. The organization of the paper is illustrated in Figure 3.

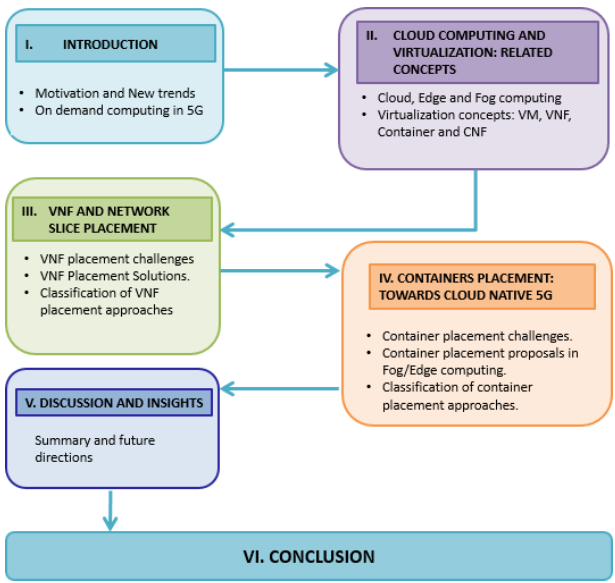

Figure 3: Pictorial view of this paper

## II. CLOUD COMPUTING AND VIRTUALIZATION: RELATED CONCEPTS

The main idea of IoT is that everything can be connected to the Internet at any time where a glut of objects (e.g., smart cameras, wearable devices, environmental sensors, home appliances, and vehicles) are connected and produce massive volumes of data. These data may be collected, integrated, processed, and analyzed to create smart cities, infrastructures, and services that improve people's quality of life. Existing IoT designs are highly centralized, relying primarily on moving data analytics, processing, and decision-making to cloud solutions. However, latency, network traffic management, computational processing, and power consumption can all be affected by data management and processing in the cloud. Moreover, in many applications that need low latency, such as health monitoring and emergency response services, the delay created by transmitting data to the cloud and subsequently back to the application can significantly influence the system's performance. Data fusion, data trends, and various decision-making approaches allow data processing closer to where data is generated and help to minimize the quantity of data transferred to the cloud, reducing network traffic, bandwidth, and energy consumption. In addition, smart cities applications such as smart health, security, and traffic control will benefit from a more agile response that is closer to real-time. Therefore, this section contrasts the cloud computing paradigm with the more sophisticated paradigms used to bring compute, storage, and control capabilities closer to where data is generated in the IoT (i.e., fog and edge computing). Also, it provides a summary of crucial virtualization concepts (i.e., VM, VNF, container).

### A. Cloud, Edge and Fog computing

*1) Cloud Computing:* Nowadays, the term cloud computing is already widespread. With the pandemic and the growth of remote working, companies have been forced to look for this type of solution to stay competitive.

IT industries have defined cloud computing from different business perspectives but the most commonly accepted definition among experts is the one provided by the National Institute for Standards and Technology (NIST) that considers cloud computing as *"a model for enabling ubiquitous, convenient, on-demand network access to a shared pool of configurable computing resources (e.g., networks, servers, storage, applications, and services) that can be rapidly provisioned and released with minimal management effort or service provider interaction"* [34]. According to IBM, *"Cloud computing is on-demand access, via the internet, to computing resources, applications, servers (physical servers and virtual servers), data storage, development tools, networking capabilities, and more hosted at a remote data center managed by a cloud services provider"* [63], cloud computing converts the IT infrastructure into a utility that provides a dynamic and scalable service-oriented IT architecture. Put simply, cloud computing includes both applications provided as services over the Internet and the datacenter hardware and software that deliver those services. Cloud computing is a global concept of anything that requires the provision of hosted services over the Internet. Infrastructure as a service (IaaS), platform as a service (PaaS), and software as a service (SaaS) are the three primary types of these services [116] as shown in Figure 4. According to NIST [128], there are five enabling characteristics of cloud computing: on-demand self-service, elasticity, resource pooling, measured service, and broad network access.

Today, cloud computing is encountering increasing challenges in satisfying the stringent requirements of new IoT applications. Latency and network bandwidth are two major issues. Future IoT solutions based on AI and emerging technologies rely significantly on the cloud as it provides nearly infinite storage and computing capacity [19]. These talents are required to turn the massive volumes of data created by the IoT into intelligent knowledge and directives. However, traditional cloud computing models are reaching their limits and are unable to handle this massive amount of data. Two new paradigms have been proposed to address these weaknesses, namely fog computing and edge computing, which allow additional computational resources (such as storage, networking, and processing) to be brought closer to the network's edge.

Table I: A comparison of our work with existing surveys based on key parameters

| Ref | Type of virtual resource | | | Infrastructure | | | Classification schemes | | |
|---|---|---|---|---|---|---|---|---|---|
| | VM | VNF | Container | Cloud | Edge | Fog | Objective functions | Heuristic and meta-heuristic algorithms | Machine learning based orchestration |
| [117] | ✗ | ✓ | ✗ | ✓ | ✗ | ✗ | ✓ | ✓ | ✗ |
| [11] | ✓ | ✗ | ✗ | ✓ | ✗ | ✗ | ✓ | ✓ | ✗ |
| [82] | ✓ | ✓ | ✗ | ✓ | ✗ | ✗ | ✓ | ✓ | ✗ |
| [40] | ✗ | ✓ | ✗ | ✓ | ✗ | ✗ | ✓ | ✓ | ✓ |
| [99] | ✗ | ✗ | ✓ | ✓ | ✓ | ✗ | ✓ | ✓ | ✓ |
| [28] | ✓ | ✓ | ✗ | ✓ | ✓ | ✗ | ✓ | ✓ | ✗ |
| [127] | ✓ | ✓ | ✗ | ✓ | ✓ | ✓ | ✓ | ✓ | ✓ |
| This survey | ✗ | ✓ | ✓ | ✓ | ✓ | ✓ | ✓ | ✓ | ✓ |

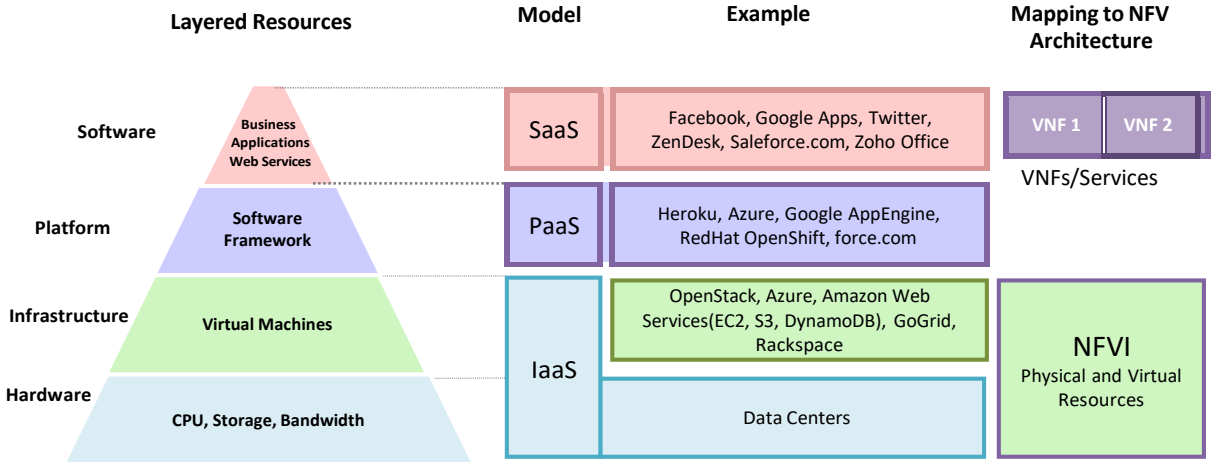

Figure 4: Cloud computing service models and their mapping to part of the NFV reference architecture

*2) Fog Computing:* CISCO introduced the concept of fog computing in 2012 to expand cloud capabilities closer to the network's edge *"Fog computing is a highly virtualized platform that provides compute, storage and networking services between end devices and traditional cloud computing data centers, typically, but not exclusively located at the edge of the network"* [26]. Since then, other definitions have emerged under various circumstances and setting. The fog is a layer that stands between the edge and the cloud, bringing the cloud closer to the IoT data processing nodes resulting in a cloud-to-things continuum that reduces latency and network bottlenecks while maintaining data privacy.

In [135], fog computing is considered as *"a paradigm to complement the cloud for decentralizing the concentration of computing resources (for example, servers, storage, applications, and services) in data centers toward consumers in order to improve service quality and user experience."*

According to NIST, *"Fog computing is a layered model for enabling ubiquitous access to a shared continuum of scalable computing resources. The model facilitates the deployment of distributed, latency-aware applications and services, and consists of fog nodes (physical or virtual), residing between smart end-devices and centralized (cloud) services."* [67].

Fog computing differs from the conventional computing models by the following features: (1) geographical dispersal, (2) contextual location awareness and low latency, (3) heterogeneity, (4) interoperability and federation, (5) real-time interactions, and (6) federated fog cluster scalability and agility [51]. In addition to these six fundamental qualities, fog computing is frequently related to (7) Wireless access predominance and (8) mobility support.

*3) Edge Computing:* Some literature considers edge computing as a synonym of fog computing [56] [55], but there are some substantial differences. In [121], edge computing refers to *"enabling technologies allowing computation to be performed at the edge of the network, on downstream data on behalf of cloud services and upstream data on behalf of IoT services"*. Edge computing concentrates on the things aspect, while fog computing concentrates more on the infrastructure aspect. The objective of edge computing is to move specific computing resources from the cloud to heterogeneous devices at the network's edge [8]. According to CISCO [62], edge computing simply refers to the concept of moving computational resources closer to data-generating devices, whereas fog computing refers to the physical implementation and management of this architecture at the cloud's edge.

Fog computing uses a multi-layered architecture to supply hardware and software operations, allowing dynamic re-configurations for diverse applications while performing intelligent activities. Edge computing provides a direct delivery service by running specific applications at a fixed logical location [32]. Edge computing tends to be restricted to a small number of peripheral devices (e.g., BS, home gateways, edge routers), whereas fog computing is hierarchical.

The fog and edge computing architectures enable the computing and storage capabilities of the network infrastructure to be leveraged for the deployment of IoT services, thereby making these services closer to the end-users. However, the network devices are heterogeneous with low computational capacity, covering a wide geographical area, and have to address the mobility of IoT users. In this way, the problem of virtual resource placement becomes more complex in terms of optimizing various parameters such as minimizing energy consumption, enhancing IoT QoS, reducing traffic congestion, and decreasing cost. MEC provides cloud computing capabilities to content providers.

### B. Virtualization concepts: VM, VNF, Container and CNF

The way network services are delivered to end-users has been transformed by NFV. Individual network services are now provided as software-based virtualized entities known as VNFs, which are dissociated from costly and specialized middle-boxes [79]. It is a piece of software that handles network tasks, including routing, switching, firewalling, and load balancing. It also eliminates the need for separate proprietary and specialized hardware from vendors, allowing network services to be executed on generic or Commercial-Off-The-Shelf (COTS) hardware with varying degrees of computing, storage, memory, and network interfaces. VNFs can be hosted using two virtualization technologies, VMs and containers.

Although VNFs are part of a conventional network design, they still have constraints as digital telecom providers progress toward offering more flexible services. When switching from physical components to VNFs, providers merely uninstalled the embedded software systems from the devices and established a large virtual machine. However, without efficiently optimizing and placing these virtual resources, this can create inefficient single-use appliances and even impact the quality of service [23].

Furthermore, the weight of VMs may restrict VNF efficiency for large-scale 5G or edge deployments that require agility, scalability, and minimal overhead. Therefore, telecom operators tend towards adopting a cloud-native approach using distributed and centralized locations that help to ensure efficiency, scalability, and reliability.

The key component of the cloud-native approach is the usage of containers rather than VMs. Containers enable users to bundle software (for example, apps, functions, or micro-services) with all of the files required to operate it while sharing access to the operating system and other server resources. This method allows the enclosed component to be moved across environments (development, test, production, etc.) and even between clouds while maintaining performance. Table II shows a comparison between VMs and containers.

Table II: Comparison between VM and Container

| Features | Virtual Machine | Container |
|---|---|---|
| Performance | Suffer from a low overhead as the instructions from the Guest to the Host OS are translated | Provide close-to-native performance compared to the Host OS. |
| Startup time | It takes several minutes for VMs to boot up. | Can boot containers in a few milli-seconds. |
| Storage | VMs require even more space as a whole OS kernel as it has to install and run the related programs. | Containers take lower space, as the basic operating system is shared. |
| Isolation | Hardware isolation. | Operating system isolation |
| Kernel | Each VM run its own OS | Containers share the same kernel |
| Benefits | Fully isolated and more secure | Lightweight, Native performance, less memory requirement, more portable |
| Drawbacks | Heavyweight, limited performance, Large memory requirement, less portable | Higher fault domain and are less secure |

As summarized in Table II, containers are lightweight alternatives to VM-based hypervisors [154] and are characterized by OS-level virtualization. In containers, a physical server is virtualized to enable autonomous applications and services to be deployed on a remote server. Unlike their VM-based counterparts, containers do not require hardware indirection and run more efficiently on the host operating system, allowing for greater application density.

Cloud-native networking functions (CNFs) are an extension of VNFs that are intended and constructed to run in containers [120]. This containerization of network architectural components enables a range of services to run on the same cluster and provides easier integration of already deconstructed applications while dynamically routing network traffic to the appropriate pods. CNFs can address some primary constraints of VNFs by shifting many of these functions into containers. Containerizing network components allows administrators to control how and where functions are executed across clusters.

## III. VNF AND NETWORK SLICE PLACEMENT

Nowadays, the telecom world is evolving exponentially to become purely virtualized. Therefore, virtualization is considered as the key element of 5G using SDN, NFV, and MEC technologies to build virtual partitioning of mobile radio access, virtual core network, and network slicing.

A VNF is an implementation of a network function that can be a firewall, a router, a load balancer, or even a mobile core network component. VNFs, in 5G network architecture (see Figure 6 ), provide complete core network functions as Home Subscriber Server (HSS), Mobility Management Entity (MME), Access and Mobility Management Function (AMF),

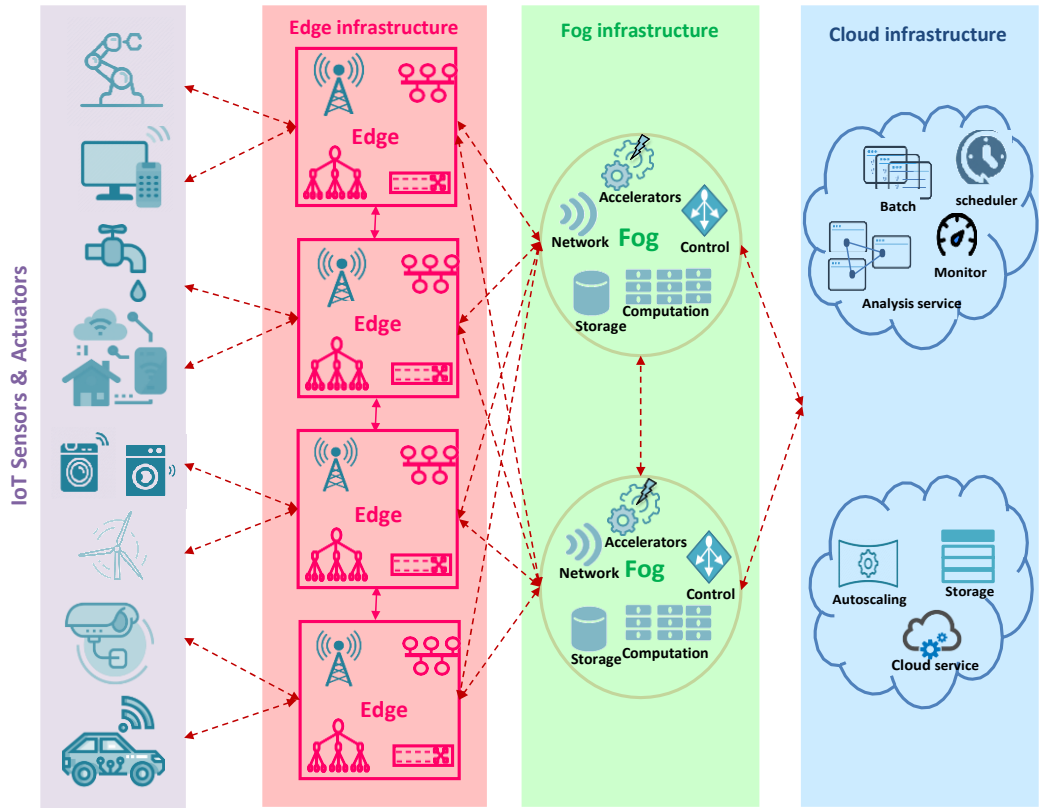

Figure 5: Cloud, Fog and Edge computing

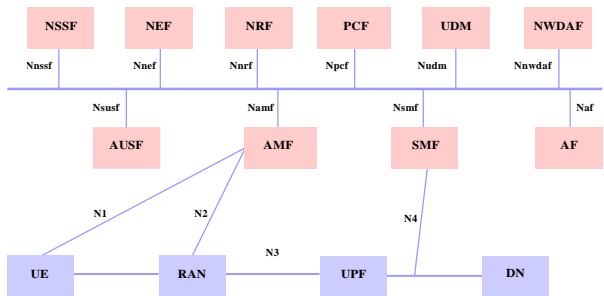

Figure 6: 5G core architecture

Session Management Function (SMF), and Policy Control Function (PCF), etc.

Regarding VM placement problems already discussed in previous surveys [20], [134], the open question is how telecom providers will work around the issues of placing VNFs in the 5G network, and the challenging task relies on where, when, and how to place the VNFs.

In this context, network slicing offers several significant advantages, which are valuable for the design of next-generation networks [45]. Slicing provides an agile VNF placement, improving network performance and decreasing operating costs. It involves deploying multiple logical networks as separate business transactions on a shared physical infrastructure [137]. Various architectures have already been proposed to provide evolved 5G infrastructures, [98], [137], [69] which offer the capabilities to support the required diversity of services, scalable deployments and network partitioning.

In [98], a 5G-ready architecture model and an NFV-based network slicing are presented to provide scalable VNFs and deliver 5G slices that meet customer requirements. In the same way, authors in [69] offer a new architecture for open cloud-based 5G Communication that treats the network slicing as a brain wave in the cloud-based Radio Access Network (RAN), aiming to increase the scalability of current RAN systems.

In network slicing, each slice has its own envelope that is a compromise related to the target usage, and its characteristics should be tailored to the chosen environment. For example, in a single 5G system, the network slicing technology can provide connectivity for smart counters using a network slice that connects IoT devices to a data service with high availability and reliability, with a given latency, throughput, and security level. At the same time, it can provide another network slice with very high throughput and low latency for an augmented reality service.

Therefore, 5G has a flexible structure where network slices assign capacity, velocity, and coverage resources separately. In

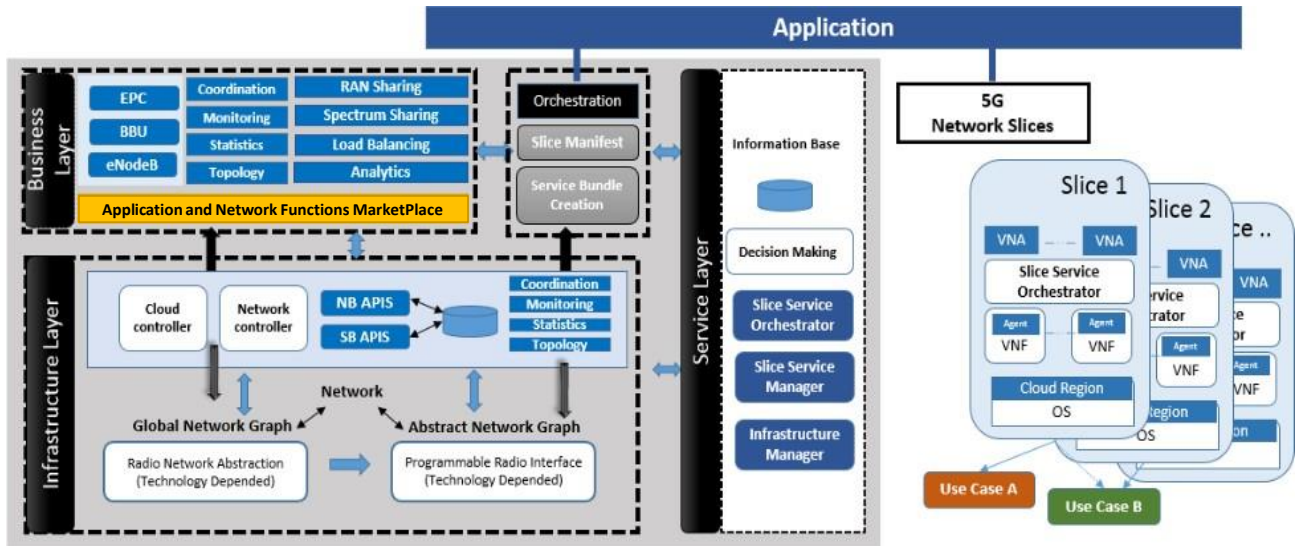

Figure 7: 5G Network Slicing architecture ( [98], [69])

this way, network slicing allows the coexistence of multiple vertical services over the same physical infrastructure. Based on [98], and [69], the network slices architecture is divided into three layers as illustrated in Figure 7:

**The business layer** is a market of applications and network functions used to provide various scenarios with different features (e.g., high mobility, speed, IoT). It creates a slice that encrypts all information required from the service layer to provide the requested function.

**The service layer** manages, configures, and scales the operational set of services according to their particular use case qualifications defined in the "slice manifest".

**The infrastructure layer** manages the re-configurable green cloud system in real-time and applies virtualization for high-speed services. The slicing in 5G typically drives two new insights, i.e., the service layer and the network slice orchestration, to supervise the life cycle of slices.

The slice orchestration is a complicated matter that can be divided into intra-slice and inter-slice problems. One of the intra-slice orchestration's crucial characteristics is the efficient placement of VNFs, including initial placement (static) and online placement (dynamic) throughout slice run-time. An intelligent placement may reduce latency, operating costs, and increase resource utilization and network performance.

Network slicing is a collection of interconnected VNFs and physical functions over a common multi-domain infrastructure to support a specific service. Its performance depends directly on the efficient placement of VNFs. For example, slices lacking low latency have to be placed close to end-users. Therefore, considering the end-to-end performance of a specific network service, VNFs must be placed in the best locations. We present in Figure 8, a scenario of VNF placement in 5G distributed edge cloud where NS is defined as a collection of VNFs required to deploy a complete 5G mobile service.

Network slices can cross various network domains, including access, core, and transport. In this context, extensive efforts have been performed to address the problems of functional placement. This section describes the VNF placement issues and challenges in terms of energy, power consumption, capacity, latency, and security. Considering the advantage of the NFV's ability to place VNFs anywhere and anytime easily, several VNF placement strategies are proposed for different NFV orchestration settings. We also present a classification of VNF placement solutions in 5G based on their objective functions.

### A. VNF placement challenges

The VNF placement requires multi-objective functions such as reducing cost, minimizing the end-to-end latency, reducing energy consumption, ensuring reliability, etc. However, the trade-offs between these objectives can lead to several conflicting issues, as placing several VNFs in the same device can cause scalability problems. For example, reducing the number of active hosts can increase network link aggregation, which affects network latency. In addition, minimizing the network latency can be impacted by VNF redundancy deployment where a cost-efficient solution is required [41]. Besides, the network power consumption must be minimized while meeting latency requirements [25]. Moreover, combining resource allocation and traffic routing raises a significant issue for VNF placement [6], where the decision to place VNFs has a critical influence on the efficient use of resources and the energy consumption in DCs. Two questions need to be addressed: how each host's computing capacity should be shared between the VNFs? and which physical machines should run the required VNFs?

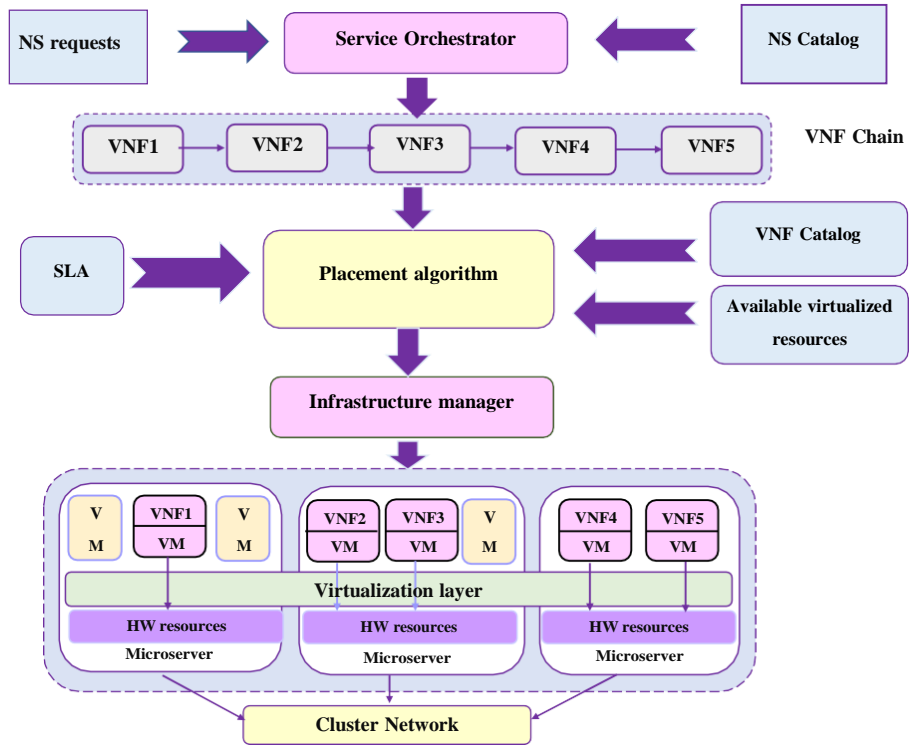

Figure 8: VNF Placement process

In order to perform services at the network edge, VMs or containers can allow VNFs to be placed on low-cost devices. When VNFs are placed near to end-users, this can minimize End-to-End (E2E) latency, response time, and even unneeded core network usage. The VNFs must be placed appropriately to handle end-user movements and address the traffic dynamics resulting in highly variable latency on network links. In addition, VNF placement in 5G networks faces substantial reliability and latency difficulties, resulting in customer dissatisfaction and revenue loss. The VNF deployed as a VM colocated with many other VMs on the same server might impact network performance when dealing with large traffic loads. Inequitable network resource sharing and VM traffic load might therefore cause increased latency.

The security risk in VNF placement is another issue that must be considered, as malware attacks can lead to substantial financial damage and loss of customers [42].

According to the literature, the different issues related to the VNF placement are energy consumption, cost, resource use, traffic, and security in order to solve the overall problem of end-to-end performance and latency (see Figure 9). VNF placement includes network functions placement, VNF forwarding graph, and VNF chains placement.

### B. VNF Placement Solutions

Mobile operators can provide specific services (e.g., social networking, video streaming, augmented reality, etc.) by chaining VNFs and routing traffic among them. Service Function Chaining (SFC) is used to configure VNFs into one logic chain

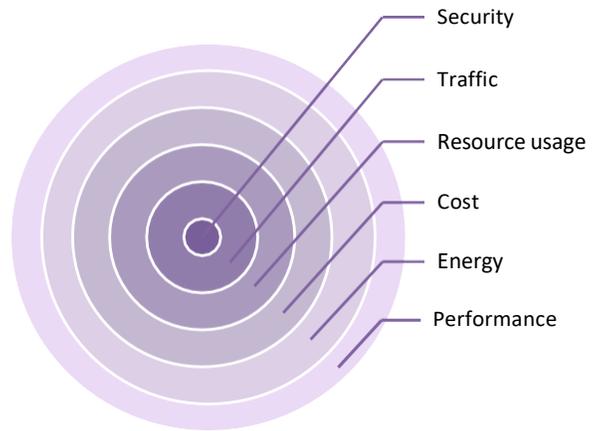

Figure 9: VNF Placement issues

with specific requirements (i.e., throughput, latency, and error rate) to deliver good QoS/QoE.

This section introduces an overall classification of the different VNF placement approaches proposed in the literature. The majority of related works handled the VNF placement problem as a multi-objective trade-off with latency. VNF placement can be carried out on different network domains: access, core, and transport. Extensive efforts have been made to solve the functional placement problem.

An optimization objective is used to measure specific as-pects of the solution generated by an algorithm. In some papers, the optimization objective may consist of a single objective, while others may be multi-objective depending on the optimization needed for the problem at hand. Generally, the more objectives used in the cost function for the considered optimization problem, the more complex the decision-making process becomes. Therefore, different trade-offs are normally set in place to balance the performance of the proposed algorithm and the quality of the generated solution. The following are considered the most common objectives utilized in the cost function definition of containers placement problem.

*1) Energy Consumption and latency:* Energy consumption is a significant concern for data centers. With the growth of network traffic, the power consumption of the infrastructure also induces a high cost for the NFV providers. Therefore, from cost control and environmental protection viewpoint, reducing power consumption is very crucial.

In [151], the authors attempt to find the optimal placement of service function chains, considering the optimization objectives for different network slices and the functional split between the central cloud and the distributed radio access point. They propose an optimization framework for placing RAN services based on an Integer Quadratically Constrained Programming (MIQCP) model and Maximum Satisfiability (MaxSAT). The problem is considered as a multi-objective approach to reduce the network latency, minimize the number of utilized nodes, reduce the power links capacities, and maximize the data throughput on the network links by keeping the bandwidth for future demands exclusively to support eMBB services. The authors analyze some scenarios using uRLLC and eMBB slices with various resource specifications. Experimental results prove that the MIQCP model is faster than MaxSAT in finding optimal solutions; however, this latter is more suitable for highly constrained problems.

[5] addresses the joint problem of VNF placement and CPU allocation decisions in the 5G network. Decisions are taken sequentially; first, the authors propose a heuristic algorithm called MaxZ that provides the deployment decisions. Then, they make CPU allocation decisions by solving the convex optimization problem of minimizing the maximum ratio latency based on the fixed results of the MaxZ heuristic. Regarding performance evaluation results, MaxZ outperforms greedy and affinity-based algorithms.

MEC in the NFV environment is considered as the 5G uRLLC enabler regarding its capacity to minimize energy consumption and end-to-end latency. The uRLLC service comprises several VNFs, where VNF placement is similar to VMP since VNFs are virtual instances performing network functions. In [145], the VNF placement accommodated for uRLLC services is formulated as an optimization approach aiming to minimize latency and maximize service availability. This model is solved using a genetic meta-heuristic algorithm. Experiment results show that this heuristic algorithm gives solutions near-optimal in less time than an exact algorithm provided by CPLEX.

In the same way, authors in [25] propose an energy-aware placement solution based on a Robust Optimization (RO) approach to minimize energy consumption while satisfying latency constraints. They use constraint modeling with Soyster heuristic model [102] to solve the problem. Their purpose consists of placing each VNF in the available network slice to the convenient VM of the service chain in a joint cloud radio architecture. The overall energy consumption is defined as the sum of energy in all assigned VNFs, and the end-to-end latency is defined as the total delay of NS, VNF, the processing delay, the path delay of NS between VNFs, and the link delay.

In [105], authors introduce a new concept of "accessible scope," defined as the group of servers used to serve a request. Instead of searching the whole servers to find the optimal placement, all servers can be divided into groups where each group can serve one specific request. The primary purpose of [105] is to minimize the server energy consumption induced in VNF placement while improving time efficiency taking into account resource constraints. Authors use the accessible scope to narrow the searching space of VNF placement and therefore reducing the searching time. They execute the Multi-Stage Graph method with the accessible scope constraint (MSGAS) to see how the size of the accessible scope affects the acceptance ratio, energy consumption, and bandwidth usage. Furthermore, the results of the algorithms with and without the accessible scope requirement demonstrate that the ones with the accessible scope constraint reduce the runtime significantly, especially for large-scale networks.

NFV can offer flexible placement of VNFs in the underlying data centers. However, the VNFs placed on the same server may experience performance interference due to shared memory and computing resources. In [94], authors propose an approach that considers energy consumption and performance interference in VNF placement. The problem is formulated as a bin-packing problem that is NP-complete. For a homogeneous environment, a First-Fit (FF) heuristic algorithm is proposed to solve the complex problem with a lower bound. For heterogeneous cases, an efficient solution named Deep Deterministic Automatic Placement (DDAP) based on Deep Reinforcement Learning (DRL) is proposed to achieve better placement. Simulation results prove that DDAP outperforms existing approaches such as FF and Ant Colony System (ACS) in terms of reducing energy consumption and running time.

In the same way, for dynamic SFC placement, two policy-based Reinforcement Learning (RL) algorithms, Proximal Policy Optimisation (PPO2) and Advantage Actor-Critic (A2C), are proposed to minimize the energy consumption while considering the availability levels required by the customer and SLA [114]. The model is formulated as a Markov decision process where SFC requests are processed sequentially. The RL algorithms yield better results than the greedy algorithm in terms of energy consumption and acceptance rate.

For the same purpose, an RL technique is used to design a VNF placement policy in an NFV architecture aiming to handle the VNF forward graph embedding problem (i.e., the resource placement in the underlying network) [126] while reducing the overall power consumption. This paper is considered an extension of Neural Combinatorial Optimization (NCO) by including constraints (e.g., SLA, latency, bandwidth, and resource utilization) in the problem definition. Simulation results prove that combining AI models with heuristic algorithms can improve the heuristic itself without requiring expertise and knowledge.

*2) Cost and Latency:* To cope with the traffic increase, mobile phone operators have to include diverse small cells by adding eNodeBs composed of indoor and outdoor types of equipment (i.e., remote radio heads (RRH) and baseband units (BBUs)) and linked via fronthaul/Backhaul to provide a 5G infrastructure called 5G crosshaul (see Figure 10 ). This new design [81] enables placing VNFs, provisioning the required network and computing resources in a flexible, cost-effective, and abstract manner.

Authors in [103] handle the VNF placement problem in service chains to ensure a reduction in operation and traffic costs. They propose an algorithm called SAMA that merges a "sample-based Markov approximation" with matching theory seeking an effective way to reduce operational and network traffic costs. This strategy first selects the nodes where VNFs can be deployed then places the VNFs in a way to minimize the total cost. Results prove the performance of SAMA in terms of reducing the cost by up to 19% compared to non-coordinated solutions.

The high cost of network power is also a significant challenge for VNF placement. In this context, authors in [89] propose a new joint placement approach for VNFs and their associated chains over the cloud computing environment. Their system performs joint node and link mapping using the extended eigen-decomposition of the request and infrastructure graphs. The main objective relies on maximum matching with minimum bipartite graph (BG) cost. This suggested algorithm achieves better performance than the greedy algorithm because it is fast stable, and its execution time depends only on NFV infrastructure size.

According to the literature, several works have addressed cost reduction by properly utilizing computing resources in cloud-based mobile core networks, seeking the optimal placement of VNFs in the same data center. For initial VNF placement, FZ. Yousaf et al. [149] propose two algorithms called Vertical Serial Deployment (VSD) and Horizontal Serial Deployment (HSD), aiming to minimize the overall cost. For highly overload profiles, the HSD can efficiently reduce the average throughput per active server since the load is shared equally across all racks with the increasing number of active servers. However, VSD leads to unbalanced distribution; servers in specific racks can be 100% utilized while other racks are underutilized or not used. To address this problem, a new automated NFV orchestrator based on machine learning [118] named zero-touch orchestration (z-TORCH) is proposed to improve the quality of management and orchestration systems by providing an optimal placement of VNFs with minimal monitoring cost.

Authors in [148] propose a dynamic placement of VNFs based on an online efficient scaling algorithm to minimize the network cost. They consider a network composed of multiple time slots with prediction stages. In each prediction stage, they apply a forecasting approach based on Fourier-Series to decide whether new demands exist in the new time slot. This online learning mechanism, based on Upper Bound Confidence (UCB), aims to reduce costs by withdrawing frequent changes in the network topology.

In [96], authors propose a heuristic approach based on bin-packing for minimizing cost in VNF placement while considering coverage, mobility, battery consumption, reliability, and low latency constraints for deploying services over a volatile 5G network. The proposed heuristic outperforms a state-of-art mobility-aware algorithm, achieving near-optimal deployments in terms of cost while enhancing convergence speed to the solution (thus increasing the number of time-feasible solutions) and reducing the number of requested handovers.

Authors in [13] handle the problem of joint traffic routing and VNF placement for a multi-cast service request to reduce both the VNF and link provisioning costs. The optimization model is formulated as a Mixed Integer Linear Programming (MILP) problem. Therefore, heuristic solutions are proposed for single path and multiple-path routing scenarios to minimize the embedding cost and provide a flexible placement and routing while ensuring low latency.

Telecom providers should handle real-time requests in the small end-to-end latency to satisfy user demands with good QoS in the 5G network; Therefore, MEC has been deployed to minimize the customer experienced delays. In [73], an SDN/NFV-enabled MEC architecture is proposed to reduce the deployment cost. However, the incurred cost for VNF placement and resource allocation (VNFPRA) in MEC nodes must be considered. The VNFPRA is formulated as a MILP problem and solved using a genetic-based heuristic algorithm to minimize the global resource cost, including the allocation cost, the computation cost, and the link usage cost. Simulation results confirm the efficiency of the suggested Genetic Algorithm (GA) based VNFPRA compared to FF and RF placement algorithms.

Similarly, in [2], authors propose a new approach for VNF placement issue for SFC using replica in the software-defined cloud, named VNF and Replica Placement (VNFRP). This approach reduces the overall SFC placement cost, service response time, energy consumption, and link bandwidth utilization. First, the problem is formulated as an ILP. Then the VNFRP heuristic algorithm is used to find the optimal placement by dynamically placing the VNFs of the SFC in the same or different nodes based on the SFC placement cost and the minimum link bandwidth.

Several approaches have been proposed to address the complexity of adjusting and placing VNFs in physical networks regarding the high number of nodes and links in DCs. The most of existing solutions focus on static placement that it initiated only if a change happened. For example, when an event occurs or some areas are busy at a certain time, it will create an overload on some servers, and therefore a VNF placement/readjustment procedure is implemented, which may cause latencies and affect the QoS. Consequently, the dynamic placement of VNFs should be deeply investigated due to the ever-changing resource availability in cloud DCs and the continuous mobility of users. The majority of VNF placement/readjustment solutions focus on optimizing objectives such as power consumption resource utilization, but they ignore important features as latency and service level objective (SLO) penalty violation cost. In this context, authors in [136] propose a Machine learning approach named MAPLE that divides the substrate network into a set of separate clusters, to reduce the complexity of VNF placement and adjustment. For the network partitioning problem, they apply the k-medoids clustering technique and a statistical technique to optimize the selection of the initial group of medoids. This helps to enhance the quality of the clusters while reducing clustering time. The VNF placement/readjustment problem is formulated as an ILP that aims to simultaneously reduce the latency, the SLO violation cost, the resource utilization, and the cost of VNF readjustment. This dynamic model helps to provide administrators with real-time placement and readjustment decisions. For large-scale DCs, they design data-driven cluster-based placement and readjustment algorithms based on machine learning that intelligently remove some cost functions from the ILP optimization problem. Simulation results prove the performance of the proposed approach in terms of reducing latency and SLO violation cost compared to existing approaches as k-means, migration without clustering, and original k-medoids.

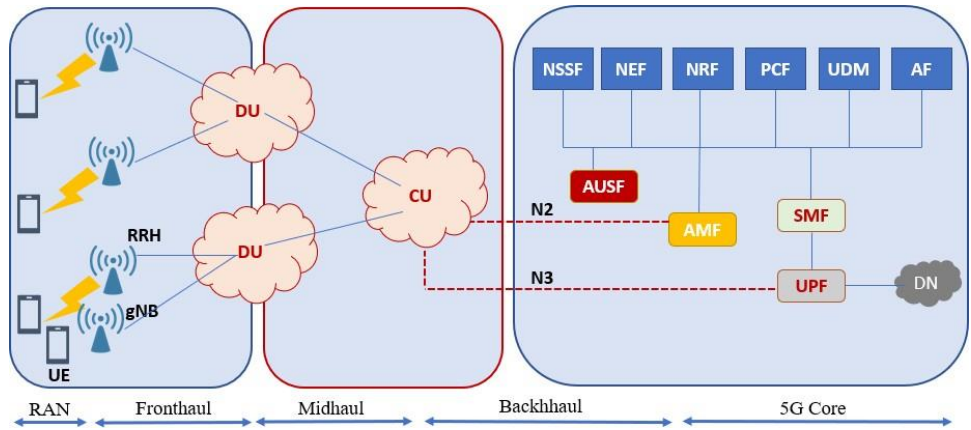

Figure 10: End to End 5G mobile network Architecture

For stateful VNFs, it is challenging to find the optimal DC for placing active and standby VNFs while reducing their overall cost, including the cost of continuous state transfer from active to standby instances, as this may result in high bandwidth consumption or even network congestion. In this respect, a RL approach is proposed for placing stateful VNFs based on a joint reservation of active and standby resources while reducing the total placement cost [70]. Simulation results prove the performance of RL based VNF placement approach in terms of improving the acceptance ratio and reducing the overall cost compared to benchmark solutions (e.g., Node-Rank [150]) in online and offline scenarios.

*3) Resource utilization and latency:* The 5G network functions are placed on VMs that can be switched between different PMs. Power consumption can be reduced by stopping unused resources. However, it is not clear what are the required resources for the network function and whether placing more VNFs in a smaller number of physical resources can degrade the service user experience and violate service level agreements.

The fast and reliable resource allocation for network slices remains challenging since each slice requires specific functionalities such as bandwidth and processing power. VNF placement and CPU allocation decisions are influenced by routing decisions from one network node to another [5]. From this point of view, remarkable effort has been dedicated for combining VNF placement, resource allocation and routing problems [76], [91], [88], [13].

The implementation of an effective framework for resource allocation to network slices remains a very relevant issue. Hence, instead of worrying about how to place VNFs individually and interconnect them, the cloud-native architecture efficiently allocates resources for network slices in terms of network bandwidth and cloud processing power [77].

In [31], the VNF placement is considered as a multi-objective optimization problem aiming to minimize the bandwidth dissipation and reduce the maximum link application simultaneously. Therefore, four genetic algorithms have been proposed using the frameworks of two existing algorithms, multiple objective genetic algorithms (MOGA) and non-dominated sorting genetic algorithm (NSGA-II): Greedy MOGA, Greedy NSGA-II, Random MOGA, and Random NSGA-II. Simulation results prove that Greedy-NSGA-II outperforms other algorithms.

In [14], the problem of VNF placement and chaining (VNF-PC) is handled by a flexible resource allocation approach aiming to minimize resource consumption in a small end- to-end delay. Authors propose a Mixed Integer Quadratically Constrained Program (MIQCP) named Flexible Resources Allocation Model (FRAM), which considers the tradeoff between resource allocation and latency, answering the question of how many resources should be allocated within the VNFs to meet the required latency. Results prove that FRAM outperforms the Strict Resource Allocation Model (SRAM) that does not consider the resource delay dependency.

In [72], a new VNF placement strategy is proposed for assigning adequate VNFs to hosts based on the total number of resources. First, before VNF placement, a periodic updating search method is applied to find the convenient host. Next, an on-demand fast VNF assignment upon request is used for placement instead of computing each time resource information.

In [22], authors propose a MILP approach to handle the joint VNF placement, resource allocation, and user association. Their optimization problem aims to reduce the service provisioning cost, minimize the effect of migration on customer's QoE, balance the resource allocation and optimize the transport network usage while guaranteeing data service requirements (e.g., latency, speed, etc.) in mobile edge computing.

The VNF Placement and Chaining Problem (VNF-PC) is one of the most challenging problems in NFV. It focuses on network resource allocation to provide end-user services such as massive IoT (mIoT). Nevertheless, these services must be supplied by an infrastructure that becomes progressively complex and heterogeneous with the growing number of network components and the exponential increase of the computational processing and runtime [110]. Therefore, as the VNF-PC is NP-hard to solve, authors in [15] first propose an ILP algorithm; however, this latter still suffers from high runtime. Second, they develop a hybrid optimization approach combining ILP with ML to minimize the number of network components and reduce the runtime in the Substrate Network (SN) through clustering strategies. The ML identifies first the patterns between requests and then decides which SN component will be used in processing. Two distinct clustering approaches are proposed: (i) based on the SN components' spatial location; and (ii) based on the SN components' historical resource usage. As a result, this hybrid strategy helps to minimize the runtime by up to 75% compared to exact methods and reduces the E2E latency without degrading the acceptance rate and provider's profit.

The challenges of network slicing and VNF placement are well debated in the literature but without considering the close relationship between the two concepts. In this context, the VNF placement over network slicing has gained attention by researchers addressing different objectives and constraints. As network slices may be implemented as a chain of VNFs, the two concepts are inextricably linked and must be explored together. The subject of resource allocation (capacity, compute, and storage) across slices has attracted much interest [54], [78], [129]. Some papers consider only spectrum resources (e.g., [78]), while others take into account also the computing resources required for VNF placement [54], [129].

In [39], the authors analyze the best VNF deployment and computational resource allocation in a hybrid two-clouds C-RAN architecture, taking into account various 5G service demands and distinctive 5G RAN functionalities. By setting limits on VNFs, the objective is to reduce the overall amount of computing resources. The problem is formulated as an ILP and solved using a standard solver. To cope with the computational cost of optimizing a large number of clouds and VNF chains, they present a simple low-complexity heuristic named Best Fit with IteRative Split Trial (B-FIRST) that tries to discover a suitable VNF placement solution with a small number of functional slices.

The NFV challenge in a C-RAN architecture is also addressed in [17], which examines six distinct criteria while formulating a C-RAN system that delivers VNFs on an edge data center. VNFs are put in the edge data center, and diverse network slices with varied requirements/constraints are considered (i.e., E2E service latency, E2E service reliability, E2E power consumption, computation capacity constraint, throughput constraint, and service admission probability). For multi-service 5G networks, authors in [122] present a new network function allocation approach that allows network functions to be deployed in a distributed computing environment based on service demands. The suggested technique includes both RAN and Core Network (CN) functions. Unlike existing systems, it provides an option capable of skewing the VNF placement based on service requirements, allowing for quick and straightforward operator-side network function deployment.

Despite the advantages of 5G networking technologies, there is a need for an automated and self-scaling orchestration system that is capable of placing VNFs dynamically to fully use MEC DCs for uRLLC services. In [36], a Deep Deterministic Policy Gradient (DDPG) RL algorithm is proposed to solve the dynamic placement of VNFs between edge and cloud network DCs. The proposed algorithm can provide the best VNF placement with respect to SLA requirement, E2E latency, and network resources compared to alternative solutions. They have proved the sustainability of DDPG for automated spatial resource allocation by migrating VNFs between cloud DCs and MEC DCs.

For large-scale networks, even the most advanced learning algorithms are unable to satisfy the complexity of VNF-FG placement. For example, the DDPG [49] is unsuitable for solving the high dimensional action space as a VNF-FG scheduling problem due to the restrictions of substrate network resources. In this regard, authors in [107] propose new approaches to find feasible solutions for the VNF-FG embedding problem by adopting the DRL technique to minimize the resource allocation while ensuring the QoS requirements. First, they propose a lightweight algorithm named Heuristic Fitting Algorithm (HFA) to deal with the efficiency of DDPG in large-scale space. The HFA determines an appropriate allocation policy of VNFs based on the proto action value received from the output of the DRL agent. Next, they propose an enhanced exploration DDPG named $E^2D^2PG$ that provides new modules in addition to HFA, i.e., evaluator and enhanced exploration, for assessing the quality of the solution and enhancing the exploration of DRL agent. Simulation results prove the performance of $E^2D^2PG$ compared to conventional DDPG and ILP. Similarly, in [106], authors suggest a DRL approach for multi-domain VNF-FG embedding. However, the results are only realized on tiny network architecture.

In [71], authors address the SFC allocation problem and provide a reinforcement learning approach for placing VNFs on an appropriate node that enhances VNF performance based on the physical network's load status. This algorithm provides good results compared to OpenDayLight (ODL) scheduler. However, this approach takes a long time to converge in a large exploration space.

In [61], an enhanced RL-based approach merged with an expert knowledge mechanism is proposed to circumvent a lengthy training procedure for VNF-FG embedding. This RL technique is based on Enhanced Q-Learning (EQL), aiming to accelerate the learning time, achieve load balancing, and improve long-term reward and performance. The EQL controls and learns the network based on the usage patterns of PMs. Simulation results, handled in large-scale networks, prove the effectiveness of EQL in terms of scalability, acceptance ratio, QoS, and acceptance gain, compared to ILP algorithms.

*4) Traffic and latency:* [76] targets the importance of combining VNF placement and path selection to maximize the served traffic demands and minimize network utilization. This problem is formulated with a mathematical program that systematically estimates a proper path length and reuses factors for each request. A usage-guided chain deployment algorithm is proposed to find a solution for optimal VNF placement in terms of reuse factor and proper path length. Simulation results prove that the proposed algorithm yields good results overcoming greedy-based and shortest-path-based heuristics. Therefore, the link capacity and resource availability should be jointly allocated in VNF placement.

In [91], the VNF placement is formulated as a mixed-integer linear programming problem. The main objective is to find an efficient placement of network functions with traffic routing among them while minimizing the CPU resource usage and the flows delay.

[88] tackles VNF placement and chaining by proposing a new analytical approach based on the Eigendecomposition method. This approach jointly manages VNF placement and traffic distribution where VNFs are placed, and traffic is spread over them all at once as tenant requests are processed collectively in the form of VNF forwarding graphs (VNF-FG). Simulation results prove that eigendecomposition based heuristic is fast, stable, and serves more requests than the greedy heuristic algorithm.

[153] also considers the VNF placement in 5G network slicing as an optimization problem aiming to achieve maximum throughput of accepted requests. A new heuristic algorithm named Adaptive Interference-Aware (AIA) is proposed to place VNFs automatically. The experimental results demonstrate the effectiveness of the proposed scheme in terms of enhancing the total throughput of slicing services such as video streaming and autonomous driving in comparison to other heuristic approaches. The AIA can also handle the traffic variations induced by VNF interference.

In [47], the authors propose a dynamic solution for joint VNF placement, traffic routing, CPU assignment, and VM activation to provide different vertical services in 5G network, considering the end-to-end delay as the primary KPI. To make this joint decision, the problem is formulated as MILP based on requests' arrival and departure times over the system's lifetime. Authors propose MaxSR, an efficient meta-heuristic method for solving the aforementioned problem for large-scale network situations based on near-future knowledge.

SDN and NFV technologies have been introduced as crucial paradigms for reaching the tactile internet's low latency requirements in multi-access edge computing (MEC) cloud systems. In [141], the authors proposed a new approach for handling distributed SFCs toward low-latency tactile internet applications, called Chain-based low Latency VNF ImplemeNtation (CALVIN). CALVIN aims to place VNFs in a distributed way with one VNF per VM. It applies fast packet input/output (IO) to prevent the metadata and batch processing of the classic Linux network stack.

Authors in [113] propose a solution for optimal VNF placement to provide eMBB services in NS by using spatial metrics of network topology. They suggest an architecture for 5G multi-tenancy networks with different software components to achieve smart decisions for VNF placement. Results prove the proposed prototype's performance in terms of computing the spatial measurements for a 5G multi-tenant network with 65538 mobile users in a small delay.

NS placement is known as an NP-hard optimization problem [43] that involves deciding which servers can host the VNFs forming the network slice and which pathways can be followed to direct traffic between these VNFs. Deep Reinforcement Learning (DRL) was recently employed in some network slice placement publications to solve this problem in a scalable fashion [146]. However, the majority of DRL research assumes a stationary environment, i.e., a static network demand. Traffic conditions in real networks are generally non-stationary and are vulnerable to large variations, such as traffic peaks, caused by unexpected events. This makes it harder for the DRL algorithm to learn the context in which slices should be placed properly. In fact, the ever-changing network environment and policies may not be in harmony with the algorithms that require previous knowledge to develop the best solutions. In [44], a new solution adapted for non-stationary traffic conditions is proposed based on hybrid DRL-heuristic algorithms to deal with the traffic change. This framework combines Advantage Actor Critic and a Graph Convolutional Network (GCN). Four algorithms have been considered pure DRL, enhanced DRL (eDRL), Heuristically Assisted DRL (HA-DRL), and HA-eDRL. Results prove that in a real non-stationary network scenario, the suggested hybrid DRL heuristic technique is more reliable than pure-DRL.

*5) QoS and Throughput:* MEC and NFV have emerged as potential technologies for delivering low-latency IoT services in smart cities. IoT devices require computing services to meet the needs of everyday applications in smart cities. Each IoT service may be implemented as a service chain made up of several interconnected VNFs running on virtual machines. When considering VNF placement, QoS is a big deal since the MEC needs to deliver high-quality IoT services. Various IoT services require different levels of QoS. For example, monitoring-related services desire a cloudlet attaining better availability as the high-availability cloudlet provides steady computing help, while game-related IoT services would like a low-latency cloudlet. As a result, an exact QoS method is required to pick appropriate cloudlets for various IoT applications. However, QoS is impacted by multiple attributes such as availability, E2E latency, resource utilization, traffic congestion, the bandwidth of communication links, etc. Thus, it is challenging to assess QoS by multiple attributes. Furthermore, the network dynamically adapts to the state of cloudlets and switches in real-time.

Determining the QoS of each cloudlet in a dynamic network becomes even more difficult. While MEC and NFV can solve the problems of resource utilization and network congestion, they also bring new challenges. To address these difficulties, authors in [157] propose a multi-attribute-based QoS approach for VNF placement and service chaining in smart cities. The optimization problem aims to maximize the throughput subject to multiple constraints such as computing resource capacity, the bandwidth of communication links, and the QoS requirement of each demand. The multi-attribute problem is formulated as an ILP, and a heuristic algorithm based on the randomized rounding technique (RRH) is proposed to solve this problem. The authors also propose an algorithm named UFPH based on an unsplittable flow approach to handle VNF placement and service chaining challenges while meeting the extra QoS requirement of each type of IoT service. Simulations results prove that the two proposed algorithms (i.e., RRH and UFPH) outperform Greedy and Random algorithms in terms of high throughput and average QoS.

*6) Security and latency:* Most previous works address different VNF placement problems without considering the resiliency of service chain embedding with the slicing concept. However, few works deal with it (e.g [144], [92]). In [144], Qi Xu et al. handle the cross-domain security problem in service function chain placement to reduce the end-to-end latency while satisfying resource constraints. This optimization problem is formulated using two ILP models for the inter and intra domains. Therefore, a heuristic algorithm is proposed to provide satisfactory solutions. In the same context, authors in [92] formulate the resiliency problem as an optimization problem aiming to reduce the maximum number of impacted service chains during a PM failure while meeting the slice-specific requirements and respecting the VNF placement constraints in the co-located network slices. Similarly, in [60], the authors provide three ILP algorithms to tackle the VNF placement issue while ensuring resiliency against single link, single node, and single-node/link failures.

### C. Classification of VNF placement approaches

In the majority of real-world scenarios, VNF placement must be treated as an online problem. On the other hand, offline methods are very important to address issues that may not be obvious in online cases, where many requests are processed in sequential order. The orchestration and placement algorithms fully understand the requirements that will be executed simultaneously in an offline manner. VNF placement algorithms are categorized into online (dynamic) and offline (static) approaches. The VNF placement problems have inspired many researchers to develop several optimization methods. These problems have been identified as NP-hard. Based on the set of papers studied in this survey, we classify the search algorithms adopted to cope with the different VNF placement problems into three types: Heuristic, Meta-heuristic (random-based) search techniques, and Machine learning algorithms, which can be described as:

· **Heuristic**: Heuristics are problem-dependent models that are designed to solve a problem according to its specification. Despite the fact that these algorithms do not always ensure convergence to an optimum solution, they are capable of obtaining competitive solutions very quickly. Table III shows a classification of heuristic algorithms based on their key parameters or objective functions.

· **Meta-heuristic**: Meta-heuristics, considered as an extension to heuristic techniques, are a high-level problem-independent algorithmic framework that can identify near-optimal solutions by iteratively optimizing solutions based on a particular performance metric. Table IV provides the meta-heuristic algorithms adopted for VNF placement issues. Meta-heuristics allow to efficiently solve the VNF placement problem. They can incorporate new objectives or constraints very easily without changing the solution, unlike heuristics. It is also important to notice that some research works demonstrate the convergence of these heuristic algorithms towards the optimum under certain conditions.

· **Machine learning** algorithms make intelligent decisions based on the data they have learned. Deep learning is a sub-field of ML that employs a layered ANN structure to learn and make smart decisions autonomously. Reinforcement learning (RL) is a field of learning in which an agent learns to make decisions through the rewards or penalties received as a result of performing one or more actions. In RL, an agent collects information about the environment, called state. Then it performs an action that moves the current setting to the next state and sends a reward to the agent. This reward reflects the measure of how the agent's action optimizes the objective function. Learning from previous experiences is a valuable capability to cope with environmental changes (e.g., change in traffic type, network configuration, etc.). Table V presents a classification of learning approaches adopted for solving complex problems in VNF placement.

## IV. CONTAINERS PLACEMENT: TOWARDS CLOUD NATIVE 5G

In the next few years, network operators will tend towards cloud architectures [66] in both edge and core network [112], to increase efficiency, reliability, and scalability. Cloud-native is the process of transitioning software deployments from traditional infrastructure to software and API-enabled infrastructure to leverage automation and DevOps techniques. This transition enhances the ability to deliver services quickly and allows providers to own their customers' experience effectively. A cloud-native strategy allows providers to deploy new services rapidly with greater flexibility. Several cloud-native principles are used to deploy 5G infrastructures, including agnosticism, application resiliency, software decomposition, orchestration, and automation. Software is divided into small components using micro-services. Each component can be individually packaged using a container as a service.

In recent years, hyper-scale clouds have evolved customer expectations around infrastructure consumption. The market has shifted to containers, micro-services and on-demand infrastructure powered by APIs and automation.

· Containers: From a basic perspective, system-level virtualization permits multiple virtual instances of an operating

Table III: Summary of papers presenting heuristic approaches for VNF placement

| Ref | Objective | Environment | Algorithms | Optimization metrics: Objectives and constraints | | | | | | Performance | Limitations |
|---|---|---|---|---|---|---|---|---|---|---|---|
| | | | | Energy | Cost | Resource | Traffic | Latency | QoS | | |
| [5] | Seek the joint optimal decision of VNF placement and CPU allocation with minimum latency and small energy consumption | SDN/NFV-based 5G network, VNF graphs | MaxZ | ✓ | | ✓ | ✓ | ✓ | ✓ | Better than greedy and affinity based algorithms | -Two stage formulation -No network cost. |
| [25] | Minimize the energy consumption while satisfying latency constraints of the NS | -MEC RAN, -NS | RO, Constraint modeling | ✓ | | ✓ | ✓ | ✓ | | Efficient placement | -No comparison with literature, -Need to consider cost. |
| [103] | Reduce operational and traffic cost | NFV Architecture | SAMA | ✓ | ✓ | ✓ | ✓ | | | Better than existing non-coordinated approaches | -Need to consider large scale network. |
| [89], [88] | Find optimal VNF placement and chaining with minimum cost and latency | Distributed cloud | Eigen-decomposition heuristic | n | ✓ | ✓ | ✓ | ✓ | | Faster than greedy | -Need to consider energy |
| [151] | Find the optimal placement of service function chains considering the multi-objective features of network slices. | -5G RAN network slicing, -VNF chains, -Distributed basebands | MIQCP | ✓ | ✓ | ✓ | ✓ | ✓ | | Better than MaxSAT | -No memory size, -Not fault tolerant |
| [149] | Reduce the deployment cost | EPCaaS | HSD, VSD | | ✓ | ✓ | ✓ | ✓ | | HSD better than VSD | -No energy consumption and QoS, -Need to consider distributed DCs. |
| [96] | Reduce cost while meeting reliability and low latency | Cloud robotics warehousing NS | Heuristic based bin-packing approach | ✓ | ✓ | ✓ | ✓ | ✓ | ✓ | Better performance than state-of-art algorithms | |
| [13] | Minimize the embedding cost for joint VNF placement and traffic routing | NFV | MILP, Heuristic algorithm | | ✓ | ✓ | ✓ | | | Efficient for large scale network | -No energy reduction, -Need to consider latency. |
| [14] | Need Reduce resource consumption in a small end to-end delay | NFV, SFC | MIQCP FRAM | | | ✓ | | ✓ | ✓ | Better than SRAM | -Need to consider large instances, -No energy, cost and traffic. |
| [22] | Address the joint VNF placement, resource allocation, and user association | MEC, SFC | MILP | | | ✓ | | ✓ | ✓ | Fastest execution time for large scale network | -No comparison with other algorithms |
| [39] | Minimize the overall amount of computing resources | C-RAN architecture | ILP, BFIRST | | | ✓ | | ✓ | | Near to optimal solution | -Need to consider realistic case, -No energy and cost |
| [153] | Enhance the total throughput of slicing services | Edge cloud, Core cloud | AIA | | | ✓ | ✓ | ✓ | ✓ | Better than SPH in terms of throughput | -Need to consider energy consumption. |
| [157] | Propose a multi attribute-based QoS approach for VNF placement and service chaining | MEC | RRH, UFPH | | | ✓ | ✓ | ✓ | ✓ | outperforms Greedy and Random algorithms | No energy and cost. |

system to run simultaneously on a single server on top of the hypervisor. On the other hand, containers are isolated and share OS kernels among all containers (see Figure 11).

Containers are widely used to optimize hardware resources, run multiple applications, and improve flexibility and productivity. Thus, a container is considered as an operating system-level virtualization technology that can be deployed on VMs or PMs and primarily used to provide a secure and isolated environment.

Table IV: Summary of papers presenting Meta-heuristic approaches for VNF placement

| Ref | Objective | Environment | Algorithms | Optimization metrics: Objectives and constraints | | | | | | Performance | Limitations |
|---|---|---|---|---|---|---|---|---|---|---|---|
| | | | | Energy | Cost | Resource | Traffic | Latency | QoS | | |
| [145] | Find the optimal VNF placement for uRLLC that reduces latency and power consumption while maximizing availability | MEC | GA | ✓ | ✓ | ✓ | | ✓ | | Better than exact algorithm provided by CPLEX | -No network traffic, -same number of vCPU. |
| [105] | Reduce the energy consumption and improve the time efficiency when placing VNFs | Large scale DC | MSGAS | ✓ | | ✓ | ✓ | ✓ | | Better than greedy and MSG | -Need to consider cost |
| [73] | Reduce the deployment cost | SDN/NFV-enabled MEC architecture | GA-based VNFPRA | | | | | | | Better than FF and RF algorithms | |
| [31] | Minimize the bandwidth dissipation and the maximum link application | NFV | Greedy NSGA-II | | | ✓ | | ✓ | | Better than NSGA-II and MOGA, Outperforms non-genetic algorithms. | -Need to consider consumed energy and cost. |
| [17] | Find the optimal VNF placement while considering six stringent constraints | C-RAN | Constraint programming | ✓ | ✓ | ✓ | ✓ | ✓ | ✓ | Close to expected 5G performance | -No comparison with other solutions |
| [76] | Maximize the served traffic demands and minimize network utilization | NFV, Edge cloud | Chain deployment algorithm | | | ✓ | ✓ | ✓ | | Better than greedy and Shortest-path (SPH) heuristics | -Need to consider online problem, -No energy consumption |
| [47] | Enhance the mobile operator profit by considering joint decisions of VNF placement, traffic routing and resource allocation | Large scale network | MILP, MaxSR | | ✓ | ✓ | ✓ | ✓ | ✓ | Better than Best-Fit | -Need to consider energy consumption. |

- Micro-services: A micro-service is an architectural and organizational pattern where every application function has its own service. These services are deployed in containers, and these containers speak with each other via APIs. The use of micro-services allows IT systems to be organized in the form of instances that can be added/removed on-demand in order to increase/decrease the scalability of their functions. However, companies that run thousands of micro-services in containers on the cloud didn't have a simple way of managing them, whereby the need for orchestration and management.

- Orchestration and automation: A few popular orchestration solutions are built to monitor the system, trigger the container's status, and balance the load between the active application instances, etc. The orchestrator used for CaaS has a direct influence on the available functions to cloud service users. Nowadays, the container virtualization market is dominated by three orchestration tools: (i) Docker Swarm multi-source cluster management and orchestration tool marketed by Docker as a native tool for managing docker clusters and container operations; (ii) Kubernetes [101], an open-source project from Google that provides a centralized system for scaling, managing containers, and automating deployment; (iii) DC/OS (the Distributed Cloud Operating System) [1], an open-source distributed operating system that enables the management of several machines in the cloud from a single interface; It allows the deployment of containers, distributed services and legacy applications in these machines and also ensures networking, service discovery, and resource management to help running and communicating services with each other.

However, regarding the benefits of orchestration in 5G cloud-native, finding the feasible placement of containers under CaaS architectures is still challenging. The container placement (CP) is similar to classical VM placement, where the main goal is to assign containers to suitable nodes to accomplish certain objective functions under specific resource constraints.

Table V: Summary of papers presenting learning approaches for VNF placement

| Ref | Objective | Environment | Algorithms | Optimization metrics: Objectives and constraints | | | | | | Performance | Limitations |
|---|---|---|---|---|---|---|---|---|---|---|---|
| | | | | Energy | Cost | Resource | Traffic | Latency | QoS | | |
| [94] | Reduce the energy consumption and performance interference | Heterogeneous | DDAP | ✓ | ✓ | ✓ | | ✓ | | Better than FFH and ACS | -No network traffic. |
| [114] | Minimize the energy consumption while achieving high availability | SFC in edge computing | PPO2, A2C | ✓ | | | | ✓ | | Better than greedy | - No network traffic, - Need to consider cost |
| [126] | Reduce the overall power consumption | NFV infrastructure | RL Advanced NCO | ✓ | | ✓ | | ✓ | ✓ | Better than FF | Need to consider cost and traffic. |
| [118] | Improve the quality of management and orchestration systems, Provide an optimal placement of VNFs with minimal monitoring cost. | Generic cloud system | zTORCH | | ✓ | ✓ | | ✓ | | Near-optimal results compared to Instant placement (i.e., without ML) | -No energy consumption, No traffic, -Need to consider geographically distributed architecture |
| [148] | Reduce the total cost while ensuring QoS and low E2E latency | NFV architecture | UCB | | ✓ | ✓ | ✓ | ✓ | ✓ | Better than greedy | -No energy consumption |
| [136] | Find real time VNF placement/readjustment solution that reduces SLO violation cost and latency | SFC | MAPLE | ✓ | ✓ | ✓ | ✓ | ✓ | ✓ | Outperforms k-means | |
| [70] | Reduce the overall cost of placing active and standby VNFs | SFC | RL | | ✓ | ✓ | ✓ | | | Better than Node-Rank | -No energy consumption and traffic. |
| [15] | Decrease the number of network components and reduce the runtime in the Substrate Network (SN) | NFV | ILP and ML | | ✓ | ✓ | | ✓ | | Better than exact algorithms | -No energy and traffic. |
| [36] | Place VNFs dynamically while considering SLA requirement, E2E latency and network resources | MEC, SDN | DDPG | | | ✓ | | ✓ | | Best placement compared to the *Baseline* | -Need to consider proactive state, -No energy and cost, -Unsuitable for large scale networks. |
| [107] | Find optimal solution of VNF-FG placement while minimizing the resource allocation and ensuring QoS requirements | NFV, Edge | DDPG-HFA, $D^3QN$ | | | ✓ | | | ✓ | $D^3QN$ outperforms FFD and DDPG-HFA | -Need to consider energy and cost. |
| [71] | Enable efficient SFC with optimal dynamic VNF placement | | RL | | ✓ | ✓ | | ✓ | | Better performance than ODL scheduler | -Long time to converge in a large space |
| [61] | Achieve load balancing and improve long term reward and performance | NFV, SFC | EQL | | ✓ | | ✓ | | ✓ | Superior than ILP | -Need to consider energy consumption. |
| [146] | Provide a NS placement solution for non-stationary traffic conditions | Large scale infrastructure, NS | HA-DRL | | ✓ | | ✓ | ✓ | | More reliable than pure DRL | |

## A. Container placement challenges

Containers are a sort of virtualization that works by separating system instances from user space inside a single

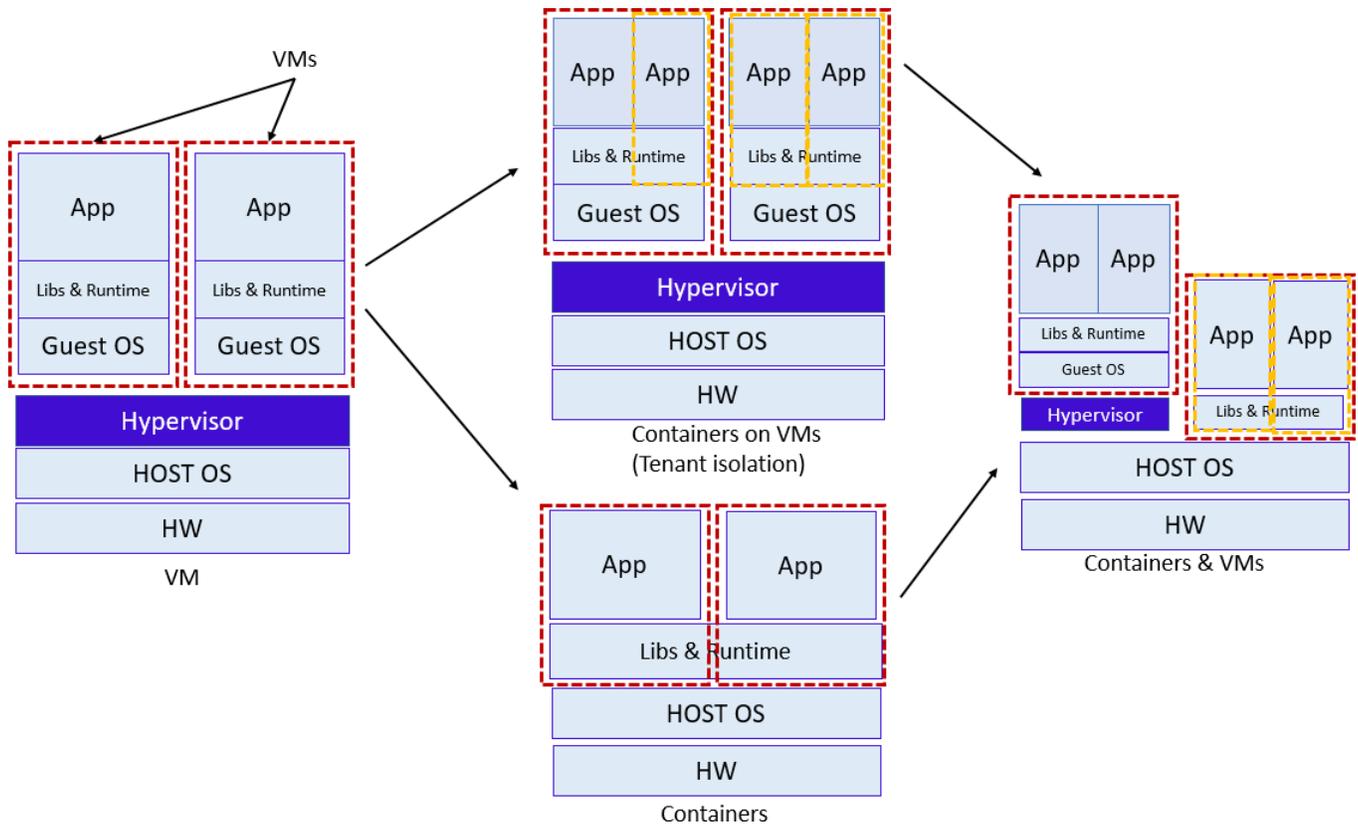

Figure 11: VM and Container deployment

(shared) OS kernel rather than virtualizing an entire machine as VMs do. Although this provides many benefits in communication performance between containers, it also means that all containers fight for the same resources in the system, leading to undesirable situations. A container is a unique process in the OS that does not have access to all of its resources. For example, it can only see a limited file system tree and cannot use all network interfaces; or it has limited memory allocation and disk I/O throughput. Current container service frameworks do not provide any kind of intelligent resource scheduling. Instead of taking a holistic view of all registered apps and available resources in the cloud, applications are often scheduled separately. This can lead to longer application execution times, resource wastage due to underutilized container instances, and a reduction in the number of apps that can be implemented considering the available resources, whereby the necessity of an optimal container placement approach that provides resource efficiency in a cloud environment [12]. In addition to resource utilization, the network traffic of containers should also be addressed to ensure the QoS and reduce the total energy consumption in a cloud environment [124]. As a result, virtual machines should be able to meet both the aggregated resource consumption and the bandwidth requirements of co-located containers. This is a challenging issue to solve due to its quadratic nature since communication between each pair of containers must be considered. Furthermore, applications should be deployed in a manner that allows them to communicate with each other with the minimum possible amount of network overhead. In this context, containers with a greater communication rate should be placed on virtual machines hosted on a single server or on servers with the shortest average network link. Network-based placement can be efficient in terms of data center power consumption and earned revenue for cloud providers.

Poor container placement may generate a bottleneck in the cloud if VMs are substantially loaded, which impacts the response time of a particular set of tasks. For example, when certain VMs are chosen to handle container loads, some of them may already be overburdened [143]. As a result, an overhead problem arises, and the response time increases. Therefore, the VMs with greater load balancing values are more convenient for placing containers. Edge computing has the potential to expand clouds by placing virtual resources (e.g., containers) closer to data sources, allowing for faster, lower-latency applications and services. Ensuring an efficient and predictable service provisioning time presents a significant and emerging difficulty as the number of Edge-servers increases and the heterogeneity of networks linking them rises. This may result in a long provisioning time depending on the container images sizes, and the network bandwidth [139]. For instance, we frequently scale-out live video stream analytics to avoid data bursts; hence, because those application's response time are in milliseconds, waiting hundreds of seconds to provision, a new container is unacceptable [37]. Therefore, container placement models are continually being advanced for edge computing to meet the KPI performance and small latencies required by IoT services.

### B. Container placement proposals in Fog/Edge computing

All previous works utilize some objectives to quantify the performance of the realized solution and evaluate the efficiency of the proposed approach to devise efficient container scheduling. Some of the well-known optimization objectives used include energy efficiency, availability, resource utilization, load balancing, scalability, cost, and makespan/latency. Other optimization objectives can be used in particular application environments or with specific data center characteristics. Readers are referred to the recent survey to examine such domain-specific optimization objectives. An optimization objective is used to measure special aspects of the solution generated by an algorithm. In some papers, the optimization objective may consist of a single objective, while others may be multi-objective, relying on the optimization required for the problem at hand. Generally, the more objectives used in the cost function for the considered optimization problem, the more complex the decision-making process becomes. Therefore, different trade-offs are typically set in place to balance the performance of the proposed algorithm and the quality of the generated solution. The following are considered the most common objectives utilized in the cost function definition of containers placement problem.

*1) Resource Utilization:* For resource management in the containerized cloud, new deployment models, such as fog and advanced mobile computing, have been established to make the cloud closer to the end-user [142] and services closer to the edge [152]. However, resource allocation in dynamic fog computing systems is a challenge. Authors in [93] proposed a joint optimization problem for container placement and task provisioning. Their main objective is to optimize the number of served end-users while considering resource utilization and mobility under delay/threshold constraints. The issue is formulated as an ILP and was solved using low complex Particle-Swarm-Optimization (PSO) based meta-heuristic and Greedy heuristic algorithms. Simulation scenarios prove that the PSO-based algorithm performs near-optimal results with more sustained execution times than the Greedy Algorithm. Besides, network slicing has been introduced by 3GPP to improve the scalability of fog computing in 5G, where the E2E network in a vertical slice connected the core network to the edge devices throughout the fog nodes. Each fog node can use the harvested energy to provide pervasive computing resources anytime and anywhere. For scalable fog computing with energy harvesting, a dynamic network slicing architecture is proposed to manage the workload handled by several fog nodes located close to each other [142] and maximize the utilization of available resources.

Previous works have separately considered the placement of VMs on PMs or the placement of containers on VMs/PMs. However, this leads to over-utilized or underutilized VMs/PMs [85]. For this reason, there is growing interest in developing a container placement algorithm that considers the simultaneous use of instantiated VMs and used PMs. Cloud-native principles and technology have proven to be an effective acceleration technology in continuously building and operating the largest clouds in the world. This new technology has been selected to develop next-generation VNFs called cloud-native Network Functions (CNFs), where network function is deployed to operate inside containers. Services are instantiated as a group of containers, which frequently leads to a high communications workload causing a degraded quality of service. Placing containers of the same service within the same server can reduce communication costs but may cause heavily imbalanced resource utilization. In this context, two phases are handled, container placement (CP) and container reassignment (CR) [84]. For the CP problem, the Worst Fit Decreasing (WFD) algorithm is proposed to provide efficient communications. For the CR problem, a reassignment algorithm named Sweap&Search is suggested to coordinate containers' distribution by migrating them among servers.

The Docker Swarm placement approach matches the containers to the available resources according to the round-robin principle without considering resource utilization of VMs or PMs. In this context, [65] proposes placing new containers on VMs while simultaneously taking into account the VM placement on PMs. The primary purpose is to reduce the number of active PMs and VMs and optimize CPU and memory usage. Therefore, the authors propose a meta-heuristic placement algorithm based on Ant Colony Optimization Best Fit (ACO-BF), which uses a fitness function that computes the percentage of wasted remaining resources in PMs and VMs. Simulation results prove the ACO-BF's performance in terms of resource usage in both PMs and VMs compared to FF and MF.

In the same way, authors in [104] have proposed a new Docker container orchestrator named Carvela. Unlike other container placement or orchestration approaches dedicated for centralized architectures [119], Carvela uses a fully decentralized architecture and resource discovery to handle a large number of volunteer resources and avoid bottlenecks; it also employs workload placement heuristic algorithms to take the appropriate placement decision with respect to the resource (i.e., CPU, RAM) and satisfying low latency and cheap bandwidth.

Sagar et al. [16] propose a dynamic resource allocation and placement algorithm (DRAP) aiming to design and place a simple cloud-native network service. Their approach can help service providers to reduce their infrastructure costs. The proposed DRAP heuristic algorithm aims to reduce resource utilization while ensuring service availability. It focuses on minimizing the number of nodes needed to place CNF pods (i.e., Kubernetes pod) by adapting the vCPU allocation to each pod. The scalable vCPU allocation permits the algorithm to scale up or down the number of pods based on service availability.

Authors in [100] present a comprehensive study of container placement algorithms and scheduling models in Edge Computing. The container placement is a decision-making problem that can be formulated by graph models or multi-objective optimization models to be solved by heuristic or meta-heuristic algorithms. In [10], the container placement problem in cyber-physical systems is formulated as an ILP optimization model aiming to maximize resource utilization while ensuring high QoS. A heuristic algorithm based on Deep Learning Artificial Neural Network (ANN) is proposed to solve the ILP optimization model.

Some scheduling methods can place containers on the infrastructure with manual resource allocation that may affect the application's performance. Therefore, an automatic approach for allocating optimal CPU resources can help to improve the efficiency of containers placement. In [4], authors propose a new deep learning-based algorithm for dynamic CPU resource allocation while reducing the job completion time. This approach employs the law of diminishing marginal returns to estimate the ideal number of CPU pins for containers in order to maximize the number of concurrent jobs while maximizing performance. Experiment results, tested on a Docker-based containerized infrastructure with real workloads, prove the performance of the proposed DL algorithm in terms of decreasing the job completion time by 23% to 74% compared to static scheduling methods as First come First Serve (FCFS), Shortest Job First (SJF), Longest Job First (LJF), and Simulated Annealing (SA).

Similarly, the Docker Swarm's scheduler overlooks the resource utilization when placing containers in the cluster. In [33], authors first examine the performance interference in container placement where results show that the performance of distributed applications can be degraded when co-located containers highly consume resources. Then, they propose a new scheduler based on machine learning clustering algorithms, □_□□□□⊕₊placement policy and doubling placement policy, that help to enhance performance while maintaining high resource consumption. Simulation results prove that the proposed placement strategies can improve the distributed application's performance by up to 14,5% compared to Random and Bin-packing algorithms.

*2) Energy Consumption:* In [155], the authors study the CP problem in terms of energy consumption; they develop a target chromosome model for optimizing energy efficiency and propose an Improved Genetic Algorithm (IGA) to find the efficient CP solution. Experiments prove that the proposed strategy is better than existing Docker Swarm strategies. Simulation results show the effectiveness and performance of IGA in terms of energy saving compared to basic GA, First-fit, and PSO algorithms.

In [58], authors propose GenPack, a new generational scheduler, for placing containers in a cloud DC to maximize energy efficiency. It learned the attributes and requirements from the system containers' runtime monitoring. Their adopted method, tested in a Docker Swarm environment, can increase energy efficiency by up to 23% compared to the built-in schedulers (i.e., Spread, binpack and random).

Container orchestration tools have emerged as an alternative to avoid the challenging problems of highly volatile workload applications and the constraints of small energy consumption and latency, though using heuristic and AI algorithms to fit the dynamic environment. In [132], a new framework called COSCO (Coupled Simulation and Container Orchestration) is developed to achieve the efficient placement of containers in fog computing environments. Besides, the authors have proposed a Gradient-based Optimization policy using backpropagation with respect to Input called GOBI to provide fast and scalable scheduling. They have also created an extended version named GOBI* to achieve QoS by providing intelligent predictions and scheduling decisions with low latency. Simulation results prove the performance of the proposed approaches (i.e., GOBI and GOBI*) in terms of reducing energy consumption, scheduling time, response time, and service level objective compared to heuristics (e.g., GA) and other learning algorithms presented in the literature.

The container placement problem can be framed into two phases (i.e., placing containers on VMs and placing VMs on PMs) while minimizing energy consumption and maximizing resource utilization. The complexity occurs when considering the heterogeneity of containers, VMs, and PMs. Instead of handling each placement separately, authors in [9] have proposed a Whale Optimization Algorithm (WOA) to solve these two placement steps as one optimization problem. The proposed algorithm efficiently minimized the overhead of creating VMs and the energy consumption compared to DGWO, TMPSO, FFD, LF, and MF.

Similarly, in [156], the initial container placement is formulated as a bi-objective optimization model aiming to minimize the power consumption while maintaining the best service performance by proposing a novel application isolation metric to quantify the overall service performance. R. Zhang et al. [156] propose an optimization approach called First Fit based on improving the Genetic Algorithm (FF-based-IGA) to find the efficient initial container placement solution. Simulation results prove that the proposed algorithm yields better results in terms of minimizing the energy consumption compared to conventional algorithms such as BF and FF.

In [90], a new scheduling approach based on a multicriteria decision algorithm is proposed to place containers in the convenient node. This approach considers three criteria: the amount of available memory, the number of containers in each node, and the number of available CPUs. The scheduling strategy aims to select the node that hosts a container by combining the Spread and the Bin Packing models to form the Technique for the Order of Prioritisation by Similarity to Ideal Solution (TOPSIS) algorithm. Simulation results prove the performance of TOPSIS in terms of reducing energy and computing time compared to Random, Spread, and Bin-packing algorithms.

Likewise, to reduce the energy consumption of service placement in cloud DC, a green container-based service aggregation is presented [95], allowing a large number of servers to be in the idle mode without impacting the quality of experience. The problem has been formulated as an optimization approach to minimize the total energy consumption with respect to service execution time. In this way, the authors propose an online learning-based technique based on Bayesian Optimization (BO) that can handle measurement noises encountered during workload characterization for containerized services. This algorithm is named Energy-Aware Service consolidation using baYesian optimization (EASY). The experiments executed in the docker swarm environment prove the effectiveness of EASY in reducing the total energy consumption and bandwidth overhead compared to FFD and BF. However, as the reduction of active nodes makes them heavy loaded, the average response time is greater than baseline methods.

The goal of resource allocation in container-based clouds is to reduce total energy consumption by properly assigning resources (such as CPU and memory) to applications without overloading the PMs. Hence, regarding the elastic nature of containers, a cloud provider must distribute appropriate resources as soon as a new request arises, and this is named online Resource Allocation in Container-based clouds (RAC) problem. It is challenging because of its two-phase strategy (i.e., placing containers in VMs and placing VMs in PMs). Previous studies learn a one-stage allocation policy for allocating containers to VMs. In contrast, the assignment of VMs to PMs is manually performed. In [130], authors present a novel Cooperative Coevolutionary Genetic Programming (CCGP) hyper-heuristic algorithm to address the RAC problem by learning the workload pattern and generating allocation rules for the two levels. Simulation results prove the performance of CCGP rules in terms of improving energy efficiency compared to the sub&Just-Fit/FF rule.

*3) Network traffic:* The containers implemented in an application are located in several PMs to ensure high parallel performance. The CNF placement has a significant impact on the network traffic and the containerized data center performance. Unlike the existing CNF placement solutions that do not consider the traffic pattern of containers, authors in [139] propose a new placement approach based on network traffic correlation named "Blender" considering the traffic between containers as a Zipf distribution. The blender approach offers two valuable benefits: (i) it reduces inter-block traffic by placing containers that often communicate in the same block. (ii) it performs efficient load balancing by grouping blocks depending on the types of required resources and dispatching them over several PMs. Simulation results prove the high performance of the Blender solution compared to SBP and CA-WFD in terms of reducing communication traffic.

The critical challenge in container cluster (CC) provisioning is the efficient placement of containers while considering inter-container traffic. This challenge is further complicated when the clusters of containers are provisioned online. Hence, authors in [158] propose an online placement algorithm to dynamically assign the container to a zone with free capacity while considering the inter-container traffic. This online placement design involves a one-shot algorithm that identifies the optimal placement for the current CC and an online algorithm framework that makes on-spot decisions upon the arrival of CC requests based on resource prices. An exhaustive sampling and ST rounding techniques were applied to reduce the complexity degree of the one-shot CC placement problem and find efficient solutions. In addition, compact-exponential and primal-dual online methods are exploited to ensure a good competitive ratio.

Monitoring inter-application traffic properly without instrumenting the application, required to dynamically determine the appropriate container placement, is difficult to achieve. In [97], authors propose an effective black-box monitoring strategy for identifying and constructing a weighted communication graph of cooperating processes in a distributed system that can be accessed for a variety of reasons, including adaptive placement.

In [138], authors propose an Availability-assured Buffered-layer Prioritized scheduler (ABP) to minimize network traffic and reduce the latency of scaling services in Docker Swarm. They adopt a heuristic algorithm named Dominant Resource Fairness to run this scheduler. Experiments show that the ABP scheduler significantly improves service creation and deployment in the Docker swarm environment.

Distributed cloud is a vital technology for 5G networks and is emerging as an alternative for managing latency-sensitive and traffic-intensive applications. Placing containers on the edge cloud enables applications to be located closer to end-users and traffic sources, which will result in reducing latency and network traffic. In this context, authors in [68] propose a two-level approach to tackle the traffic and latency-aware container placement optimization problem in a distributed cloud. They use an ILP model to solve the first step of placing containers in DCs, considering the average cost, resource utilization, and acceptance ratio as key performance metrics. For the second step of placing containers in servers, a traffic-aware heuristic algorithm is proposed. Results prove the performance of the proposed heuristic approach in reducing all traffic metrics compared to conventional bin-packing heuristics (i.e., FFD, BFD).

*1) Response time, Execution time, communication cost:* Nowadays, telecom operators tend to deploy their 5G services in the form of containers in their large-scale data centers. Each service includes multiple modules that are instantiated as a group of containers, where containers owned by the same service commonly need to communicate with each other to provide the required service [86] leading to cumbersome inter-server communication and service performance degradation. The communication cost can be significantly decreased if these containers are placed on the same server. Nevertheless, containers belonging to the same service are typically exhaustive on the same resource. For example, containers of data transfer applications are network I/O intensive [87], thus resulting in unbalanced resource usage, but this can have a positive impact on availability, response time, and system throughput. However, reducing the communication cost while maintaining a balanced use of resources is challenging. In this context, for this conflicted goal, authors in [84] handle the problem in two phases: container placement and container reassignment. The first one aims to place a set of containers on DCs to minimize resource utilization while reducing the communication cost by using a Worst Fit Decreasing (WFD) algorithm. The second phase of container reassignment attempts to optimize a given container placement by migrating containers between servers. The authors have proposed a two-stage approach, Sweep & Search, that first tackles overloaded servers and then optimizes the targets using local search techniques. Simulation results prove the performance of the proposed algorithms in terms of low cost, high throughput, and balanced resource utilization compared to the state-of-art algorithms.

The optimal container placement in volatile fog nodes

(i.e., CPU, communication fabric, memory, storage) can be ensured by reducing costs and guaranteeing the customer's QoS (i.e., response time and isolation). In [80], a two-phase partition-based optimization approach is proposed to improve the service availability and QoS satisfaction through initially matching applications to fog device communities and then placing application services transitively on fog devices. However, in this case, the inter-container communication was neglected, and few are the papers [84], [27] that addressed the impact of network communication on isolation and QoS in fog computing. In [27], authors have proposed an optimal genetic algorithm for container placement aiming to reduce the response time while considering the heterogeneity of fog nodes and inter-container network communication as well as the isolation requirements for applications deployed on fog computing networks. For inter-container communication, three modes have been deployed ( i.e., host mode, overlay mode, and RDMA). Results prove the performance of the proposed GA algorithm compared to greedy and ILP in terms of isolation and significant response time reduction when using a greater number of RDMA-enabled fog nodes.

The convenient placement of containers on VMs can help to optimize resource utilization in cloud environments. However, the bad placement may result in a bottleneck in the cloud if VMs are highly congested, which can adversely impact the response time of a given set of tasks. In [46], authors propose an Ant Colony Optimization (ACO) algorithm to reduce the overall makespan of tasks, thus leading to reduce the response time of applications. The typical ACO tends to schedule tasks to the most used node, which can cause an overload issue if the node is carrying a large load. The drawback of ACO is its disregard of resource utilization and energy efficiency. Considering these challenges, authors in [52] have proposed a Modified ACO (MACO) for container placement to optimize the response time and improve the scheduling decision while taking into account resource utilization, throughput, and energy consumption. Simulation results show that MACO outperforms the First Come First Serve algorithm (FCFS) in response time and throughput.

Similarly, authors in [125] have proposed a container-based task scheduling using a hybrid bacteria foraging optimization (HBFA) algorithm to minimize the execution time and increase the resource utilization in an edge computing environment. The proposed HBFA yields better scheduling results than BF and GA.

Current serverless platforms have multiple constraints in terms of supporting data-centric distributed computing [], which are compounded by the operational features of underlying edge systems, particularly when it comes to function placement decisions. Among constraints, the high latencies incurred by the distance between nodes in edge computing infrastructures. Therefore, the inter-node proximity and bandwidth must be taken into account []. In [109], a new container scheduling system called Skippy is proposed to provide an efficient placement of edge functions by considering the scheduling limitations of the application's data flow, network topology, proximity, and available compute capabilities. Experiments prove the performance of Skippy in terms of reducing execution time, cloud execution cost, and uplink usage.

5G networks, driven by NFV, promise to enable a wide range of services from various market segments (e.g., Smart Cities, smart homes, Automotive, etc.). Services must be connected in a precise order to properly benefit from NFV, forming a Service Function Chain (SFC). The majority of existing works handle the placement of VNF-based service chains, with the target to find the optimal placement while minimizing the end-to-end latency and maximizing the resource utilization [14]. Many optimization models based on ILP [57], [3] have been proposed to facilitate the SFC orchestration through deciding whether to migrate or replace VNFs while reducing the SFC latencies. Nonetheless, few papers considered the latency-aware container-based SFC chain in fog computing. In [115], authors propose an SFC controller, as an extension of Kubernetes scheduling features, to optimize the placement of container-based SFC while optimizing resource provisioning and minimizing the E2E latency.

Several placement techniques based on deep reinforcement learning (DRL) have been proposed in cloud or edge computing environments, but they are not suitable for distributed architectures. The task of forming efficient DRL agents involves a lot of training data, and their procurement is expensive. The centralized DRL-based strategies suffer from poor scalability and are therefore unable to solve placement issues. Many IoT applications are created using Directed Acyclic Graphs (DAGs) with different topologies. Meeting the requirements of DAG-based IoT applications leads to further constraints and makes the placement problem more complex. To address these issues, authors in [48] propose a distributed DRL approach named X-DDRL that aims to solve the placement challenge of DAG-based IoT applications while minimizing the energy consumption and the execution time. For training the distributed brokers, an actor-critic-based distributed application placement technique named IMPALA (IMPortance weighted Actor-Learner Architectures) is proposed to achieve timely and efficient application placement decisions. IMPALA framework can minimize the agent's exploration costs and provides rapid convergence to optimal solutions.

*2) QoS:* In smart applications (e.g., smart cities and smart homes), the big data workflow is based on various sensors and video streams where AI and feature extraction techniques are performed. The captured information is stored in DB containers. These containers need to be placed on Edge, Fog, or Cloud infrastructures while addressing the QoS requirements. Open source solutions such as Docker or Kubernetes can orchestrate containers in edge and fog computing, but the decision on where to place a software instance considering major QoS metrics (i.e., throughput, latency, power consumption, CPU utilization, cost). In [74], a stochastic approach for DB container placement based on Markov Decision Process (MDP) is proposed to (i) dynamically enhance the automation based on new QoS attributes, (ii) build utility functions that provide the reward values and help to find the optimal solution of decision making, (iii) ranking deployment infrastructures based on rewards to get the QoS success score. The authors also propose a new architecture that automates the whole process. The author's experiments were based on 25 infrastructures and 8 QoS attributes. Simulations prove that MDB is better than Analytic Hierarchy Process (AHP) method in terms of QoS violations, where no violation is faced for MDB, in contrast to AHF, where QoS violation is omnipresent in all workload scenarios.

In a multi-level environment of cloud/fog/edge, the network has a crucial role as it serves as a communication link between all of the system's participants; as a result, its performance has an impact on the whole system. Therefore, the network should be considered while making all placement decisions in order to meet the QoS performance. In [53], the QoS assurance techniques in fog computing are categorized into service/resource management, communication management, and application management. Accordingly, the container QoS can be satisfied by managing network and storage workloads. In [24], authors propose a container management strategy named CONtrol to balance the bandwidth between storage traffic and application traffic over a hyper-converged architecture. The primary objective of CONtrol is to manage the storage traffic when scheduling containers while sustaining the QoS of the container network. CONtrol aims to make dynamic placement decisions over bandwidth redistribution across diverse workloads using a proportional-integral-derivative controller. In the same way, a new scheduler module (i.e., an extension of Kubernetes scheduler) is proposed to make placement decisions based on network status [30]. Here, the authors develop a network-aware scheduling algorithm named IPerf aiming to compute the estimation time for job completion where the system rejects the applications that do not meet the deadline.

*3) Security:* The majority of container placement research focuses on dealing with resource utilization, energy consumption, cost, response time, etc. However, few works consider the security problems. In this subsection, we cite some security strategies treated in previous works, such as avoiding co-location attacks, improving user isolation, and identifying vulnerabilities. The various co-resident attacks present a significant challenge to cloud providers and tenants. Each type of co-resident attack needs significant changes in hardware, host systems, container engines, and system configurations. On the other hand, as the overall co-residence is unknown (and increasing), it is difficult to fix the software and hardware to address future attacks. The container deployment approach offers a straightforward and effective way to influence the likelihood of co-residency.

In [123], authors confirm that containers placed in VMs are susceptible to co-residency attacks. The co-residency detection can tolerate background noise with a 70% success rate, as long as it does not surpass the hardware capacity. Their analyses show that any change in architecture and orchestrator can reduce detection fidelity by up to 10%. Therefore, cloud customers should not rely on orchestration platforms to satisfy sufficient protection against co-residency attacks.

In [75], authors propose a Secure Container Deployment Strategy named SecCDS based on Genetic Algorithm (GA) to cope with co-resident attacks in container clouds. They carefully orchestrate the placement and migration of containers to dissociate attackers and victims on various PMs. The GA-based strategy aims to overcome the problem of selecting the target to which the containers migrate. Meanwhile, to increase the convergence time of GA, a Simulated Annealing (SA) algorithm is proposed by performing a strong neighborhood search for each unit in GA. Simulations prove that the proposed approach yields better results in terms of minimizing the co-residency attacks with negligible effect on system performance and workload compared to classic strategies such as Previous Selected Server First(PSSF), Random, Most and Least strategies.

### C. Classification of container placement approaches

The container placement approaches can be categorized into two types: queuing and concurrent placement.

The queuing approach can be defined as a FIFO or priority-based approach where the CP decision is performed on a container by container [133], [83], [108], [29]. The container-by-container placement approach has the benefit of making parallel decisions on a distributed architecture; however, it also has critical restrictions for ensuring efficient placements of all concurrent tasks as it does not have a holistic view of pending containers. In general, an optimal decision is difficult to achieve since, in queuing models, the initial placement decision is taken by the first container in the queue regardless of the other remaining containers in the queue. However, if the specifications of all requests are known beforehand, some scheduling rules for smart container placement can be developed based on machine learning algorithms to predict all incoming requests [131], [38].

The concurrent approach is defined as a batch processing concept where computing requests are first gathered, and then a placement decision is made [64], [18]. Here, the scheduler has a complete view of all workloads, where all containers can be placed in convenient VMs. However, the concurrent scheduling strategy may be highly complex, as the problem is often formulated as integer programming or mixed-integer programming, which may impact the QoS. Also, the batching time needed for intermittent requests can delay the placement as the allocation task waits for a certain time to serve multiple requests.

Figures 12 and 13 depict an example of the two approaches (i.e., queuing and concurrent) for four concurrent requests. As depicted in Figure 12, for queuing model, container four can not be placed on any node due to the lack of resources after placing the three other containers. On the other hand, as seen in Figure 13, the container scheduling is optimized when using the concurrent approach.

The strategies used for container placement are: Spread (tries to place the containers evenly on available nodes), Binpack (place the containers on the most-loaded host that still has enough resources to run the given containers), and custom. Similar to VNF placement, the surveyed scheduling algorithms adopted for container placement are classified into three categories: Heuristics, meta-heuristics, and machine learning.

The majority of the reviewed techniques use some heuristics to get approximate solutions to the problem, as shown in table

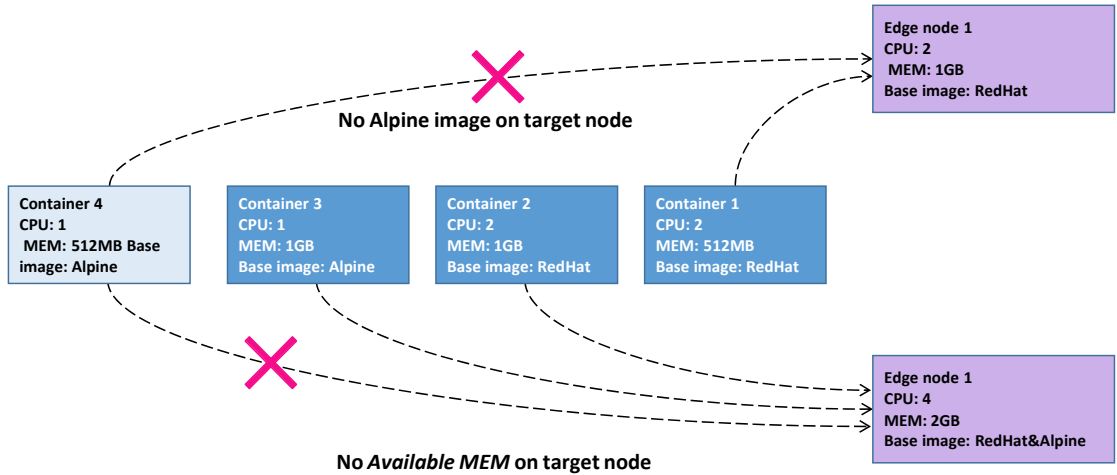

Figure 12: Queuing approach

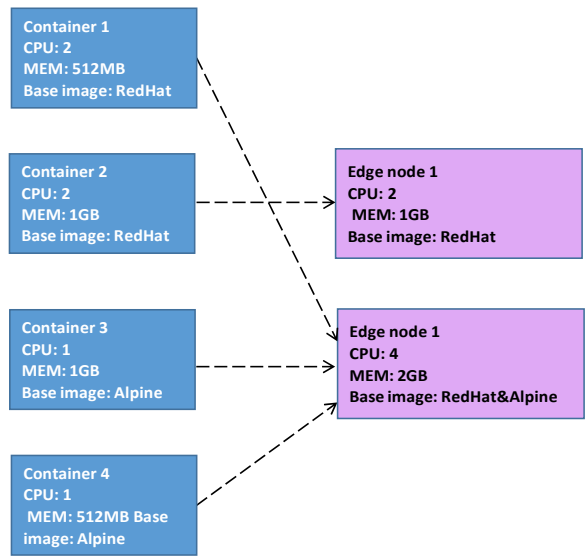

Figure 13: Concurrent approach

VI. Heuristic algorithms are typically of low complexity and generate a suitable program in a reasonable time.

Meta-heuristics are a flexible and popular class of population-based optimization algorithms inspired by the intelligent processes and behaviors of nature. Meta-heuristics are widely used to solve optimization problems in several disciplines. Two important features of these algorithms are a selection of the fittest and adaptability to the environment. Table VII provides a classification of meta-heuristic algorithms used for container placement.

Machine learning is an active field of research with a lot of success in various applications, and it is very promising for container placement. ML algorithms are successful because of the availability of big data to train the model. Compared with other heuristics, one can benefit from machine learning techniques to improve solution accuracy and effectiveness by making intelligent scheduling decisions. We present in Table VIII a classification based on performance metrics and machine learning algorithms.

## V. DISCUSSION AND INSIGHTS

For a telecom operator, providers are converging towards containerized architectures. Therefore, instead of focusing on the placement of VMs or VNFs, the placement and orchestration of containers in edge and fog computing must be taken into account to provide services with good performance and high QoS.

For VM, VNF, or container placement, the optimization approaches were classified into six objective functions, where minimizing energy consumption is the most interesting one regarding its huge impact. As a synthesis of this paper, a placement strategy has the following objectives:

· Minimize energy consumption.
· Reduce the number of active networking elements.
· Maximize resource utilization.
· Reduce the network traffic.
· Increase load balancing.
· Minimize SLA violations.
· Reduce cost.
· Reduce latency, response time, execution time.
· Increase the Return on investment (ROI).
· Ensure high performance.
· Guarantee good Quality of service (QoS).
· Reduce co-residency attacks.
· Increase security.

However, it is a bit difficult to optimize these several conflicting objectives simultaneously. The VNF/container placement problem is defined under various parameters regarding the type of objective functions, the constraints, and the environment. These parameters differ from one scheme to another, and choosing the problem setting depends heavily on the context and the scope of dealing with virtual resource placement. For example, if high energy consumption is the most critical issue in a DC, it would be a primary objective in the problem formulation, and the other less critical issues can be ignored or considered as constraints. Therefore, based on the literature review, this mono-objective problem can be solved using heuristic or deterministic algorithms. However, if there are more than two or three conflicting goals, the problem is considered a multi-objective optimization approach. Several approaches have been proposed to solve this type of problem, based on heuristic and meta-heuristic algorithms. Besides, machine learning algorithms are well recommended for large data centers and distributed architectures to find optimal solutions for complex problems.

This paper classifies placement techniques into three categories based on the adopted optimization algorithms. We explore the optimization objectives to evaluate the performance of the generated scheduling. We have described and evaluated existing strategies for each type based on their key performance indicators to identify their advantages and limitations. The future expansion of container technology will create major changing standards, requiring the development of new orchestration solutions, placement, scheduling, and resource management. Emerging technologies like Edge/Fog computing and micro-services provide new ways to provide real-time schedulers that are sensitive to energy, communication, inter-container traffic, and security variations for these environments.

Furthermore, designing a security-conscious scheduler to prevent security threats connected with containers when deployed across cloud infrastructures might be an interesting subject of future research. The sustained success and attraction of deep learning algorithms can help to build smart scheduling policies by predicting future workloads and capacities for container placement and consolidation, load balancing, and resource provisioning. More fine-grained system counters are expected to control and gather metadata for prediction and decision making in order to make the greatest use of deep learning algorithms. Furthermore, to handle energy usage, SLAs, and QoS, multi-objective holistic management container scheduling strategies must be explored.

## VI. CONCLUSION

MEC and NFV have emerged as promising technologies to deliver low latency slicing services in the 5G communication network. However, because of the large number of nodes and links in today's Data Centers, NFV raises a number of challenges where the most significant one is the difficulty of placing VNFs in physical networks and the inter-dependency among VNFs composing a given network service. Several contributions have been made in an attempt to address these challenges in a static or dynamic manner. This paper provides a classification of the existing virtual resource (VNF or Container) placement methods and algorithms. The optimization objective may consist of single-objective or multi-objective depending on the performance metrics required for the problem at hand. We categorize the scheduling techniques into three folds: heuristics, meta-heuristics, and machine learning algorithms. Besides, we identify the performance metrics, advantages, and limitations for each category. We also highlight the convergence towards cloud-native infrastructures.

Table VI: Summary of papers presenting heuristic approaches for container placement

| Ref | Objective | Environment | Algorithms | Optimization metrics: Objectives and constraints | | | | | | Performance | Limitations |
|---|---|---|---|---|---|---|---|---|---|---|---|
| | | | | Energy | Cost | Resource | Traffic | Latency | QoS | | |
| [84] | Balance resource utilization and minimize communication cost | Large scale DCs | WFD, Sweep&Search | | ✓ | ✓ | | ✓ | | -WFD outperforms Bin-packing and Random, Sweep&Search better than Greedy | -Need to include traffic and energy consumption. |
| [104] | Handle the large number of volunteer resources and avoid bottlenecks with low latency and bandwidth | Decentralized architecture | Caravela with heuristic algorithms | ✓ | | ✓ | | ✓ | | Better than random approach | No traffic. |
| [16] | Reduce the resource utilization while ensuring the service availability | NFV | DRAP | | ✓ | ✓ | | ✓ | ✓ | Better than Gurobi algorithm | -Simple CNFs, -No energy and traffic |
| [156] | Reduce the power consumption while ensuring best service performance | NFV | FF-based-IGA | ✓ | ✓ | ✓ | | | | Better than FF and BF | -Homogeneous VMs and PMs, -Static placement, -Need to consider network traffic interference. |
| [130] | Reduce the total energy consumption by properly assigning resources to applications without overloading the PMs | Container-based clouds | CCGP | ✓ | | ✓ | | ✓ | ✓ | Better than sub&Just-Fit/FF | -Need to consider the network traffic. |
| [139] | Reduce inter-block traffic and provide efficient load balancing between resources | containerized DCs | Blender | | | ✓ | ✓ | ✓ | ✓ | Better than SBP and CA-WFD | -Need to consider migration. |
| [138] | minimize network traffic and reduce the latency of scaling services | Cloud | Dominant Resource Fairness, Abp scheduler | | | ✓ | ✓ | ✓ | ✓ | Better than default scheduler | -Need to consider dynamic placement. |
| [68] | Address the traffic and latency-aware container placement optimization problem | Distributed cloud | Traffic-aware heuristic | | ✓ | ✓ | ✓ | ✓ | ✓ | Better than FFD and BFD | -Need to consider energy consumption. |
| [80] | Improve the service availability and QoS satisfaction | Fog computing | Two-phase heuristic | | | ✓ | | | ✓ | Outperforms ILP in terms of response time | -Need to consider total service cost and network usage. |
| [125] | Reduce the execution time and increase the resource utilization | Edge cloud | HBFA | | ✓ | ✓ | | ✓ | | Better than BF and GA | -Need to consider energy consumption and resource estimation. |
| [109] | Provide an optimal placement of edge functions, by considering the scheduling limitations of application's data flow, network topology, proximity and available compute capabilities | Edge cloud | Skippy | | ✓ | ✓ | | ✓ | | | -Need to consider auto-scaling and workload migration. |
| [74] | Provide automatic container placement considering QoS metrics | Edge and Fog | MDP | | ✓ | ✓ | ✓ | ✓ | ✓ | Better than AHP | -Need to consider energy consumption and migration. |
| [30] | Enhance the QoS and reduce the time of job completion | Fog computing | IPerf | | ✓ | ✓ | ✓ | ✓ | ✓ | Better than AHP | -Need to consider energy consumption and migration. |

Table VII: Summary of papers presenting meta-heuristic approaches for container placement

| Ref | Objective | Environment | Algorithms | Optimization metrics: Objectives and constraints | | | | | | Performance | Limitations |
|---|---|---|---|---|---|---|---|---|---|---|---|
| | | | | Energy | Cost | Resource | Traffic | Latency | QoS | | |
| [93] | Optimize the number of served users while considering the resources utilization and mobility under delay/threshold constraints | Fog computing | PSO | | | ✓ | | ✓ | ✓ | Better than greedy and near to optimal | -Need to consider traffic congestion. |
| [65] | Minimize the number of active PMs and VMs and optimize resource utilization in terms of CPU and memory usage | CaaS cloud | ACO-BF | | | ✓ | | ✓ | | Better than BF and MF | -Need to consider traffic. |
| [155] | Optimize energy efficiency | CaaS | IGA | ✓ | | ✓ | | ✓ | | Better than basic GA and PSO | -Static container placement, -Need to consider network traffic and cost. |
| [9] | Reduce the energy consumption and the overhead of creating VMs | CaaS cloud | WOA | ✓ | | ✓ | | ✓ | | Better than DGWO, TMPSO, FFD, LF and MF | -Static container placement, -Need to consider network traffic, migration time and SLA. |
| [158] | Place dynamically container to a zone with free capacity while considering the inter-container traffic | Cloud container cluster | Online algorithm | | ✓ | ✓ | ✓ | ✓ | ✓ | Better than greedy and rounding algorithms | -Need to consider energy consumption. |
| [27] | Reduce the response time while considering the heterogeneity of fog nodes, inter-container network and the isolation | Fog computing | GA | | ✓ | ✓ | ✓ | ✓ | ✓ | Better than greedy and ILP | |
| [46] | reduce the overall makespan of tasks, thus leading to minimize the response time of applications | Fog computing | ACO | | | | | ✓ | ✓ | Better than random and FF | -Need to consider resource utilization and energy efficiency. |
| [52] | Optimize the response time and improve the scheduling decision while taking into account resource utilization, throughput and energy consumption | CaaS | MACO | ✓ | ✓ | ✓ | | ✓ | ✓ | Outperforms FCFS | -Need to consider SLA. |

Table VIII: Summary of papers presenting learning approaches for container placement

| Ref | Objective | Environment | Algorithms | Optimization metrics: Objectives and constraints | | | | | | Performance | Limitations |
|-----|-----------|-------------|------------|--------|------|----------|---------|---------|-----|-------------|-------------|
| | | | | Energy | Cost | Resource | Traffic | Latency | QoS | | |
| [10] | Maximize the resource utilization while ensuring high QoS | Cyber physical systems | ANN | | | ✓ | | ✓ | ✓ | Very efficient in saving resources | -No comparison with other algorithms, -Need to consider traffic management. |
| [33] | Enhance performance with high resource utilization | Docker swarm and Amazon ECS | K-means+ | | ✓ | ✓ | | ✓ | ✓ | Better than random and bin packing | -No network traffic. |
| [4] | Provide dynamic CPU resource allocation while minimizing the job completion time and increasing the applications performance | Docker-based containerized infrastructure | Deep learning | | | ✓ | | ✓ | ✓ | Better than static scheduling (i.e., FCFS, SJF, LJF, SA) | -Need to consider disk IO and memory, -No traffic. |
| [58] | Maximize energy efficiency | Docker swarm environment | GenPack | ✓ | ✓ | ✓ | | ✓ | ✓ | Better than Random, Spread and Binpack scheduling | -Need to consider network and disk intensive. |
| [132] | Provide efficient placement of containers while minimizing energy consumption, SLO violation and response time | Fog computing | GOBI, GOBI* | ✓ | ✓ | ✓ | | ✓ | ✓ | Better than GA and other learning algorithms | -Need to consider serverless computing. |
| [90] | Optimize the containers placement by minimizing memory and CPU usage and the number of nodes | Cloud platform | TOPSIS | ✓ | ✓ | ✓ | | | | Better than Random, Spread and Bin-packing | -Static placement, -No fault-tolerance. |
| [95] | Reduce the total energy consumption without impacting the QoE | Cloud | EASY | ✓ | ✓ | ✓ | | ✓ | ✓ | Better than FFD and BF | -The average response time is more than baseline -Need to consider the overload. |
| [48] | Reduce the energy consumption and the execution time | Heterogeneous fog computing | X-DDRL | ✓ | ✓ | ✓ | | ✓ | | Better than other DRL-based techniques | -Need to consider the dynamic change of transmission power. |